\documentclass[preprint,aps]{revtex4}
\newcommand{\til}{\tilde t}
\newcommand{\tom}{\tilde\omega}
\newcommand{\refeq}[1]{(\ref{#1})}
\usepackage{graphicx}
\usepackage{epstopdf}
\begin{document}
\title{Physical aging studied by a device allowing for rapid thermal equilibration}
\author{Tina Hecksher}
\email{tihe@ruc.dk}
\affiliation{DNRF Centre ``Glass and Time'', IMFUFA, Department of Sciences, Roskilde University, Postbox 260, DK-4000 Roskilde, Denmark}
\author{Niels Boye Olsen}
\email{nbo@ruc.dk}
\affiliation{DNRF Centre ``Glass and Time'', IMFUFA, Department of Sciences, Roskilde University, Postbox 260, DK-4000 Roskilde, Denmark}
\author{Kristine Niss}
\email{kniss@ruc.dk}
\affiliation{DNRF Centre ``Glass and Time'', IMFUFA, Department of Sciences, Roskilde University, Postbox 260, DK-4000 Roskilde, Denmark}
\author{Jeppe C. Dyre}
\email{dyre@ruc.dk}
\affiliation{DNRF Centre ``Glass and Time'', IMFUFA, Department of Sciences, Roskilde University, Postbox 260, DK-4000 Roskilde, Denmark}
\date{\today}

\begin{abstract}
Aging to the equilibrium liquid state of organic glasses is studied. The glasses were prepared by cooling the liquid to temperatures just below the glass transition. Aging following a temperature jump was studied by measuring the dielectric loss at a fixed frequency using a microregulator in which temperature is controlled by means of a Peltier element. Compared to conventional equipment the new device adds almost two orders of magnitude to the span of observable aging times. Data for the following five glass-forming liquids are presented: Dibutyl phthalate, diethyl phthalate,  2,3-epoxy propyl-phenyl-ether, 5-polyphenyl-ether, and triphenyl phosphite. The aging data were analyzed using the Tool-Narayanaswamy formalism.  The following features are found for all five liquids: 1) Each liquid has an ``internal clock'', a fact that is established by showing that the aging of the structure is controlled by the same material time that controls the dielectric properties. 2) There are no so-called expansion gaps between the long-time limits of the relaxation rates following up and down jumps to the same temperature. 3) At long times the structural relaxation is not stretched, but a simple exponential decay. 4) For small temperature steps the rate of the long-time exponential structural relaxation is identical to that of the long-time decay of the dipole autocorrelation function.

\end{abstract}


\maketitle

\section{Introduction}\label{intr}

The change of materials properties over time is referred to as aging. Aging phenomena often involve chemical degradation, but there are also several instances of purely physical property changes. Understanding physical aging is important for many materials applications; moreover physical aging presents fundamental scientific challenges and provides valuable insight into material properties. This paper shows that by utilizing the Peltier thermoelectric effect physical aging may be studied at considerably shorter times than has so far been possible. The new setup adds almost two decades to the span of aging times compared to what may be obtained by conventional equipment in the same observation time.
 
A prime example of aging is that of a viscous liquid's physical properties relaxing slowly when a perturbation is applied to the equilibrium liquid close to its glass transition. In equilibrium a liquid's properties do not change with time, of course (the fact that the liquids studied below are supercooled and thus technically only in quasiequilibrium is not important, because no crystallization was observed). If temperature is changed, properties gradually adjust themselves to new equilibrium values. If temperature is lowered, a glass is produced -- recall that by definition a glass is nothing but a highly viscous liquid that has not yet had time to equilibrate \cite{kau48,deb96,ang00,alb01,kob04,dyr06}. Any glass ages towards the equilibrium liquid state. This state that can only be reached on laboratory time scales, however, if the glass is kept just below the glass transition temperature -- contrary to popular myth, windows do not flow observably.

Aging is a nonlinear phenomenon. This is because the aging rate is structure dependent and itself evolves with time when the structure changes as equilibrium is gradually approached \cite{kov57,kov63,moy76,maz77,str78,sch86,mck94,hod94,hod95,avr96}. Thus aging studies provide information beyond that obtained by linear-response experiments like, e.g., dielectric relaxation measurements. There are good reasons to believe that on the microscopic level aging is heterogeneous \cite{edi00,ric02}; the below analysis is, however, entirely macroscopic and does not discuss possible microscopic interpretations of the observed aging phenomena (see, e.g., Refs. \onlinecite{die03} and \onlinecite{ric10}).

A typical aging experiment consists of a temperature step, i.e., a rapid decrease or increase of temperature to a new, constant value. Ideally, such a temperature step should be instantaneous; more precisely the new temperature should be established as constant in time and homogeneous throughout the sample before any structural relaxation has taken place. If this is achieved and if sufficient time is available, it is possible to monitor the complete relaxation to equilibrium of the physical property being probed. An experimental protocol that measures the complete relaxation curve will be referred to as an ``ideal aging experiment''\cite{kol05}.

What are the requirements for an ideal aging experiment? First, there should be good temperature control and the setup should allow for rapid thermal equilibration following a temperature jump. Secondly, a physical observable is needed that may be monitored quickly and accurately and which, preferably, changes significantly even for rather small temperature changes. The latter property allows for studying aging following temperature jumps that are of order just one percent in absolute units, which is enough for most ultraviscous liquids to become highly nonlinear. The organic liquids studied in this paper have glass transition temperatures in the region 170K-200K and most temperature jumps presented are just one or two K jumps.

\begin{figure}
\includegraphics[width=12cm]{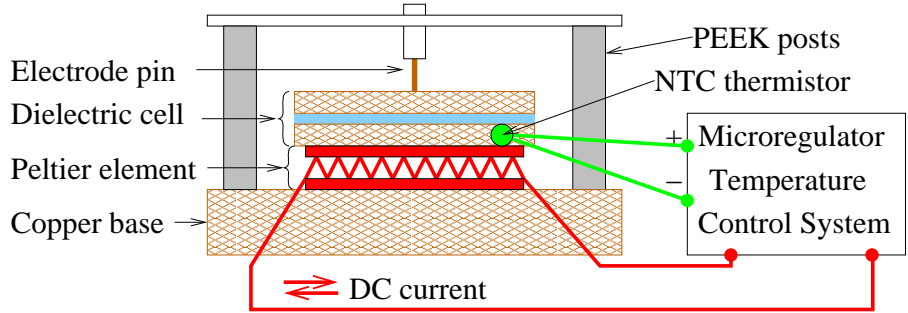}
\caption{\label{microreg}Schematic drawing of the dielectric measuring cell with the microregulator. The liquid is deposited in the (50 $\mu$m) gap between the disks of the dielectric cell. The Peltier element heats or cools the dielectric cell, depending on the direction of the electrical current powering the element. The current is controlled by an analog temperature-control system that receives temperature feedback information from an NTC thermistor embedded in one disk of the dielectric cell. A stainless steel electrode pin keeps the cell pressed against the Peltier element and provides electrical connection to one of the disks. The dielectric measuring cell is placed in the main cryostat. Details of the setup are described in Refs. \onlinecite{iga08a} and \onlinecite{iga08b}.}
\end{figure}

In current state-of-the-art aging experiments the characteristic thermal equilibration time $\tau$ is at least 100 s, if $\tau$ is defined from the long-time thermal-diffusion-limited approach to equilibrium $\sim \exp(-t/\tau)$. This reflects the fact that heat conduction is a notoriously slow process. Experience shows that in order to monitor an almost complete aging curve at least four decades of time must be covered; for instance the typical aging function $\exp(-K\sqrt{t})$ decays from 97\% to 3\% over four decades of time. Thus with present methods, one needs at least of order $100\,{\rm s}\times 10^4 = 10^6\,{\rm s}$ to have an almost ideal temperature down-jump experiment. This is more than a week. Clearly much is to be gained if it were possible to equilibrate sample temperatures faster.

In order to make possible faster temperature-jump experiments, we designed a dielectric cell based on a Peltier thermoelectric element by means of which the heat flow is controlled via electrical currents (Fig. \ref{microreg}) \cite{iga08a}. The characteristic thermal equilibration time of this ``microregulator'' is two seconds. This is almost a factor of hundred times faster than that of conventional equipment, which usually involves much larger heat diffusion lengths; our liquid layer is just 50 $\mu$m thick and the use of a Peltier element minimizes heat diffusion lengths outside of the liquid layer. In the microregulator temperature may be kept constant over weeks, keeping fluctuations below 100 $\mu$K \cite{iga08a,iga08b}. 

For monitoring aging we chose to measure the dielectric loss (the negative imaginary part of the dielectric constant) at a fixed frequency. With modern equipment this quantity may be measured quickly and accurately. Moreover, for a viscous liquid of molecules with a permanent dipole moment, a large frequency range exists in which the dielectric loss changes considerably for small temperature variations. The dielectric loss was used previously for monitoring aging by several groups, e.g., by Johari \cite{joh82}, Schlosser and Sch\"{o}nhals \cite{sch91}, Alegria {\it et al.} \cite{ale95,ale97,goi04}, Leheny {\it et al.} \cite{leh98,yar03}, Cangialosi {\it et al.} \cite{can05}, Lunkenheimer {\it et al.} \cite{lun05,lun06,weh07}, D'Angelo {\it et al.} \cite{dan07}, and Serghei and Kremer \cite{ser08}.

This paper presents ideal aging experiments on five organic liquids for temperature up and down jumps (Sec. \ref{exp}). The temperature jumps are of order one percent, and as mentioned aging is monitored by measuring the dielectric loss at a fixed frequency in the Hertz range. In Sec. \ref{internal} we give a mathematical formulation of the reduced time concept; this is sometimes referred to as the material time or -- perhaps more intuitively appealing -- the time measured on an ``internal'' clock, i.e., a clock with clock rate varying with temperature and with the annealing state of the sample. We here follow Narayanaswamy's seminal paper from 1971 \cite{nar71} and go into some detail in order to make the paper easier to read for nonexperts in aging. In Sec. \ref{howto} a new test of the existence of an internal clock is proposed. In contrast to most earlier works this test makes no assumptions regarding which quantity controls the internal clock's rate or the mathematical form of the relaxation function. This section demonstrates that all five liquids have internal clocks. Section \ref{longtime} extends the data analysis in order to study whether the long-time relaxation is stretched or simple exponential. Section \ref{calibrating} shows that within the experimental uncertainties the long-time simple exponential structural relaxation has the same rate as the long-time exponential decay of the dipole autocorrelation function. Finally, Sec. \ref{sum} gives a brief summary and a few concluding remarks. A discussion of noise and systematic errors in the data analysis is given in the Appendix.

\section{Experimental results and initial data analysis}\label{exp}

The experimental setup is detailed in Refs. \onlinecite{iga08a} and \onlinecite{iga08b}, which describe the microregulator, the surrounding cryostat, and the electronics used for measuring the frequency-dependent dielectric response. We studied aging of the following five organic liquids: Dibutyl phthalate (``DBP''), diethyl phthalate (``DEP''), 2,3-epoxy propyl-phenyl-ether (``2,3-epoxy''), 5-polyphenyl-ether (``5-PPE''), and triphenyl phosphite (``TPP''). These liquids are excellent glass formers. In order to ensure complete equilibrium before each measurement, the sample was kept at the temperature in question until there were no detectable changes of the dielectric properties.

\begin{figure}
\includegraphics[width=7.9cm]{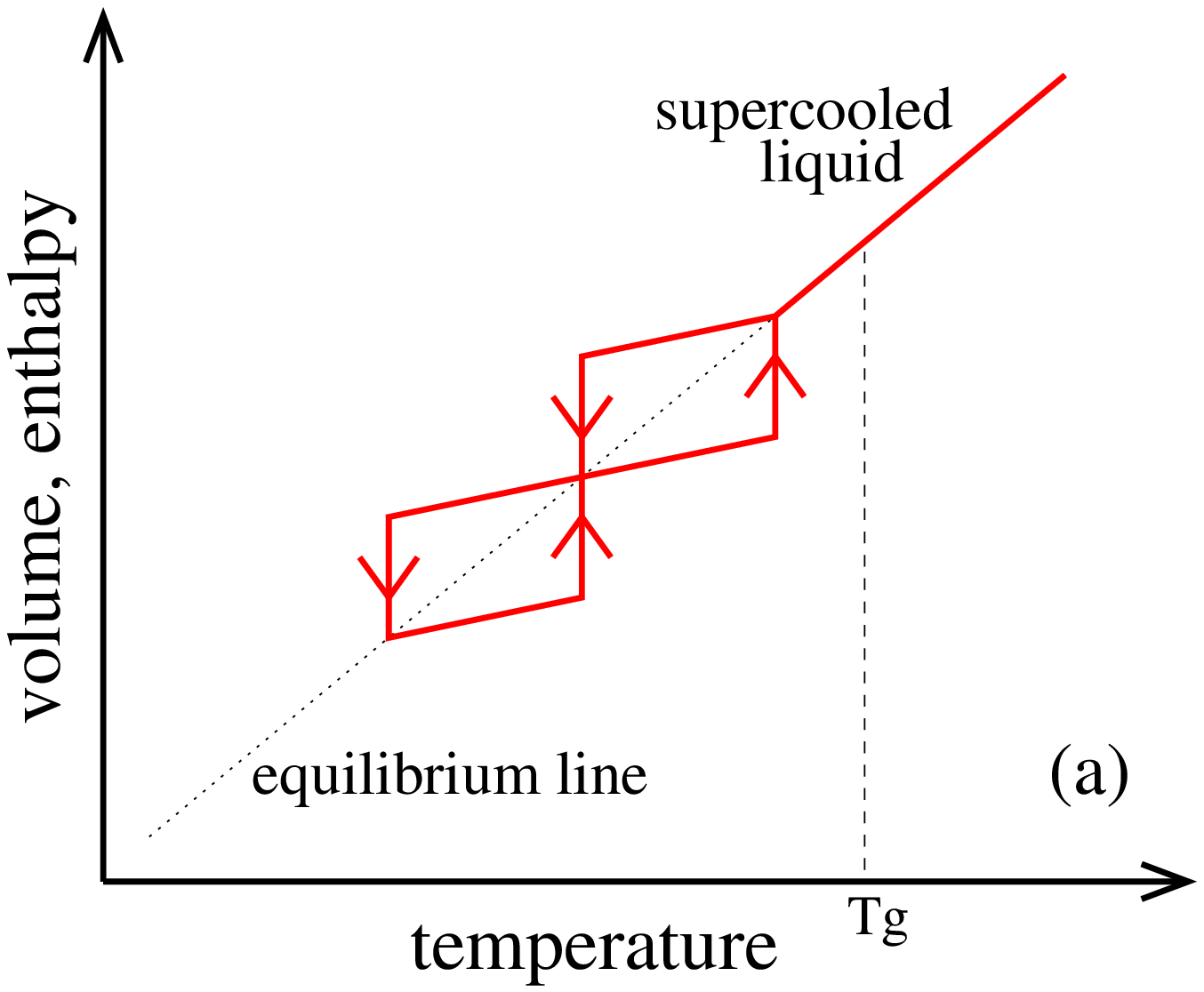}
\includegraphics[width=7.9cm]{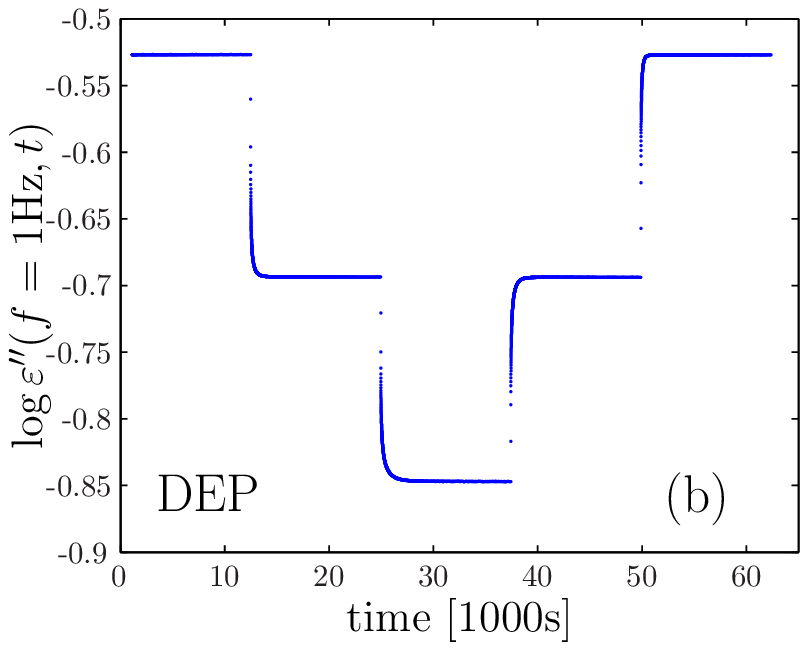}

\caption{\label{expprot}A typical measurement.
(a) Schematic representation of the protocol in which the sample is first aged to complete equilibrium to a temperature slightly below the calorimetric glass transition temperature $T_g$, a process that typically takes weeks, followed by two down temperature jumps and two up temperature jumps. 
(b) Data from measurements on DEP following this protocol, jumping from 184K to 183K, further to 182K, and back to 183K and finally to 184K. The dielectric loss $\varepsilon''$ was measured as a function of time at the frequency $f=1 {\rm Hz}$. The duration of the measurement depends on the temperature range, i.e., how long it takes to equilibrate the sample fully after a temperature jump. Following this procedure we know the relaxation functions as well as the equilibrium values of the dielectric losses at the temperatures in question.}
\end{figure}

Aging was studied by monitoring how the dielectric loss at a fixed frequency, $\varepsilon''(f)$, develops as a function of time following a temperature jump. In order to avoid the liquid aging significantly during the measurement of a single frequency response data point, the monitoring frequency $f$ must be considerably higher than the inverse structural relaxation time that is of order the inverse main (alpha) loss-peak frequency; thus the monitoring frequency must be much larger than the loss-peak frequency. For the data analysis of Secs. \ref{internal} and \ref{howto} to apply, however, $f$ must also be sufficiently below any contribution(s) from potential beta processes. These constraints vary with the liquid and the selected temperature range, and the choice of $f$ was optimized for each liquid. For all five liquids the optimal $f$ is in the Hertz range. 

Measurements consist of consecutive temperature jumps of (usually) one or two K, in most cases with two down/up jumps followed by two up/down jumps. This is illustrated in Fig. \ref{expprot}, which in (b) shows the raw data obtained for DEP. Here $f=1$Hz and the temperature jumps are one K. The temperature protocol ensures that data are obtained for one up and one down jump to the same temperature. The duration of each measurement varies with the relaxation time of the liquid in question at the measured temperatures. A time-consuming part of the experiment is the initial aging to complete equilibrium at some target temperature just below the calorimetric glass transition temperature, which in most cases required weeks of annealing. In all cases care was taken to ensure that the loss at one temperature was monitored until the sample had reached complete equilibrium; only thereafter was temperature changed to a new value. The Appendix discusses possible sources of errors in the experiments.

\begin{figure}
\includegraphics[width=7.9cm]{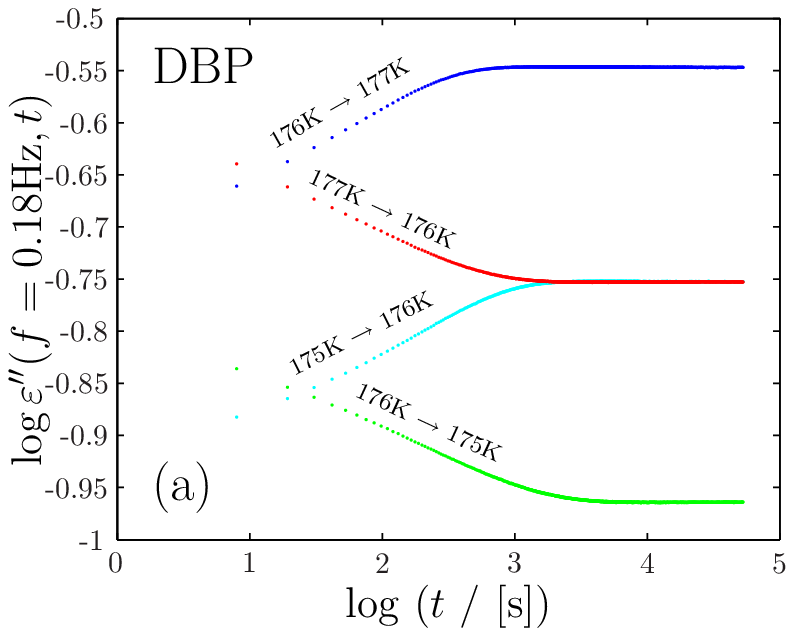}
\includegraphics[width=7.9cm]{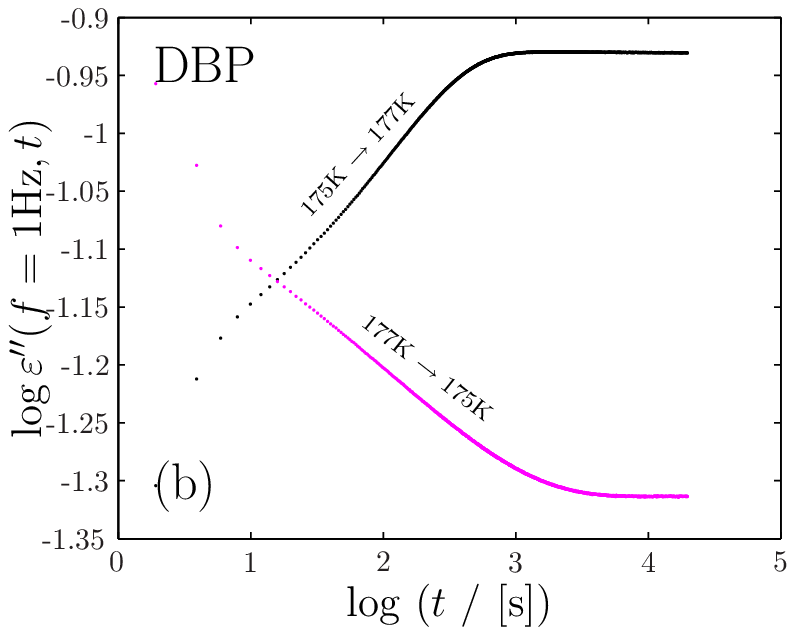}
\includegraphics[width=7.9cm]{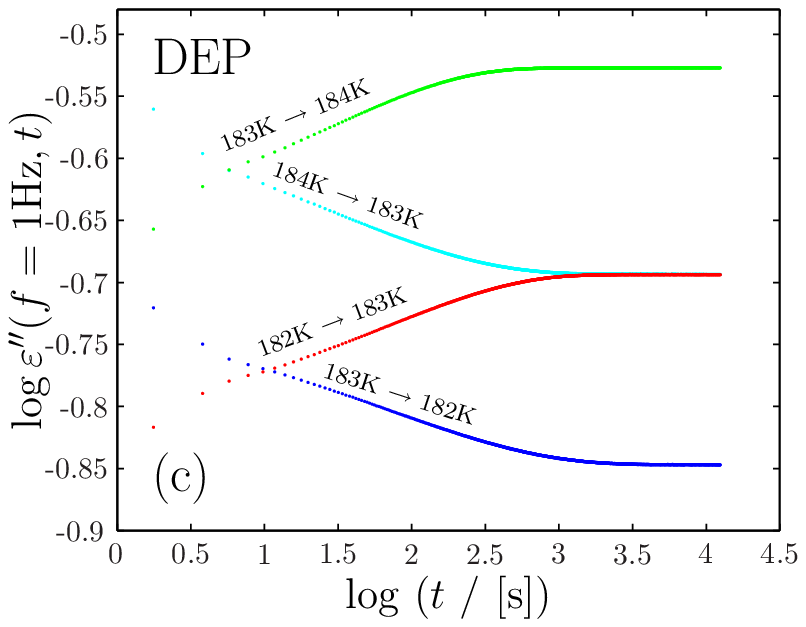}
\includegraphics[width=7.9cm]{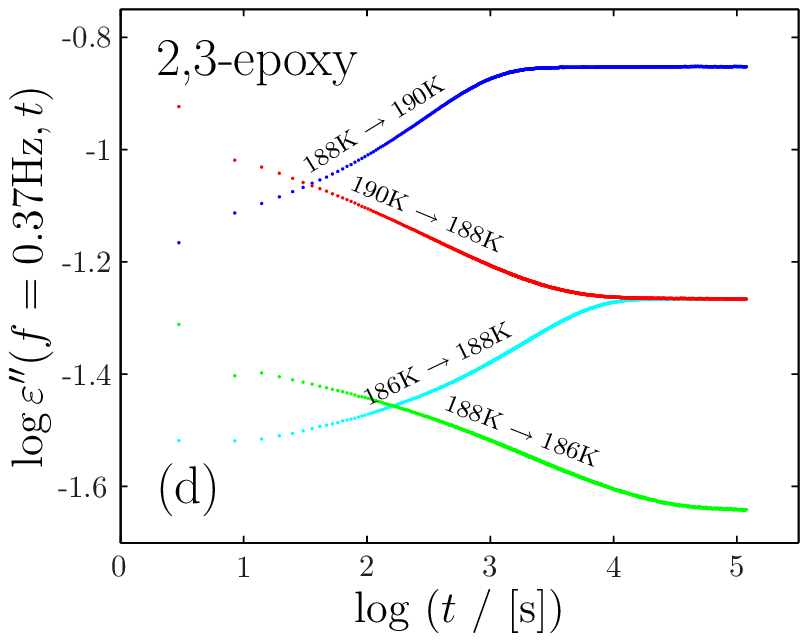}
\includegraphics[width=7.9cm]{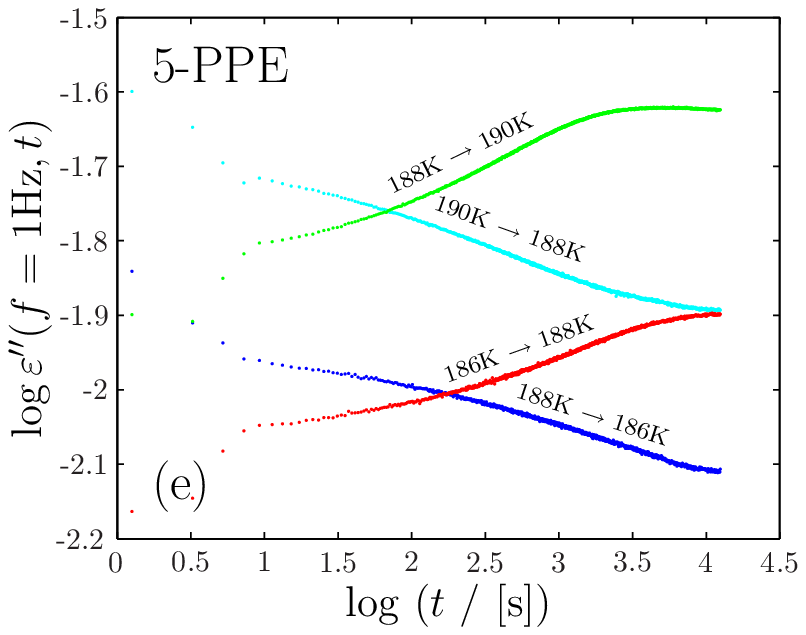}
\includegraphics[width=7.9cm]{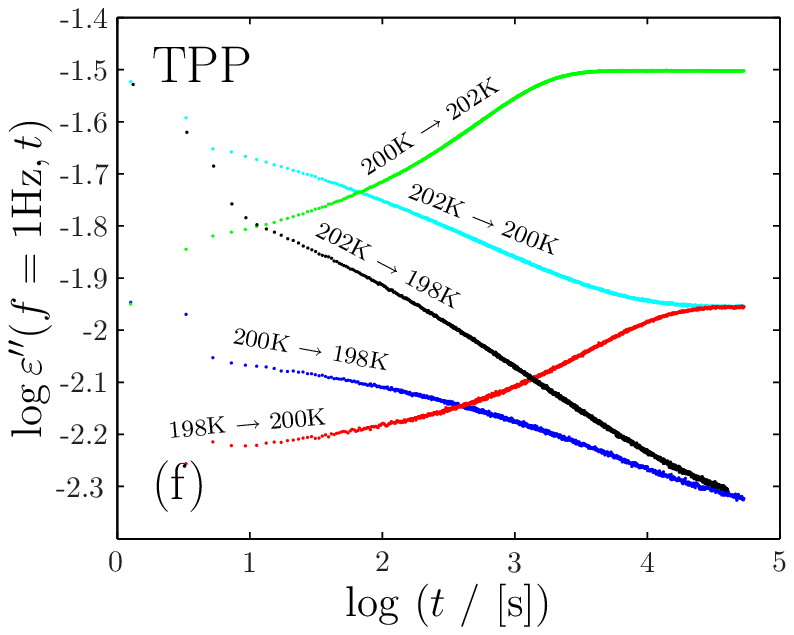}
\caption{\label{raw}Monitoring aging by measuring the dielectric loss at a fixed frequency. This figure presents the full set of data upon which the paper's analysis is based \cite{data}. The data are given in log-log (base 10) plots showing the dielectric loss as a function of time. In all cases the starting situation (``$t=0$'') is that of thermal equilibrium, a condition that is ensured by annealing for such long time that no observable change is seen in the dielectric loss. 
(a) Dibutyl phthalate (``DBP''). A series of measurements at $f=0.18$Hz stepping 1K from 177K and 175K to 176K, as well as the reverse. 
(b) DBP stepping from 175K to 177K and back, this time monitored at $f=1$Hz. 
(c) Diethyl phtalate (``DEP'') ($f=1$Hz). 
(d) 2,3-epoxy propyl-phenylether (``2,3-epoxy'') ($f=1$Hz). 
(e) 5-polyphenyl-ether (``5-PPE'') ($f=1$Hz). 
(f) Triphenyl phosphite (``TPP'') ($f=1$Hz).}
\end{figure}

Figures \ref{raw}(a)-(f) show the data upon which the paper is based. Two data sets were included for DBP, with aging  monitored at different frequencies. Note that aging for down jumps to a given temperature is faster than for an up jump ending at the same temperature (compare, e.g., the two jumps to $200$ K in Fig. \ref{raw}(f)). This is the so-called fictive-temperature effect described already by Tool in the 1940's \cite{too46}, an effect which comes from the fact that the relaxation rate is structure dependent and itself evolves with time: A down jump is ``autoretarded'' \cite{kov63} because as the structure ages, the aging rate decreases. In contrast, an up jump is ``autoaccelerated'' because as the structure ages, the aging rate increases \cite{kov63}. These are nonlinear effects that are characteristic for structural relaxation of single-component systems (but not, e.g., for aging involving composition fluctuations in binary systems). The fact that the fictive-temperature effect is clearly visible in Fig. \ref{raw} shows that even relatively small temperature jumps are highly nonlinear, reflecting the fact that the equilibrium relaxation time is strongly temperature dependent for glass-forming liquids.

For any experiment monitoring the relaxation of some quantity towards its equilibrium value, the normalized relaxation function $R(t)$ is defined by subtracting the long-time (equilibrium) limit of the quantity in question and subsequently normalizing by the overall relaxation strength \cite{sch86,nar71}. In the DEP case, for instance, where the quantity monitored is $\log\varepsilon''(f=1 {\rm Hz})$, for a temperature jump from $T_1$ to $T_2$ starting from equilibrium the normalized relaxation function is given by 

\begin{equation}\label{relax_fct}
R(t)\, = \, \frac{\log\varepsilon''( f = 1{\rm Hz},T_2,t) - \log\varepsilon''   (f=1{\rm Hz}, T_2, t\rightarrow\infty)}{\log\varepsilon''(f =  1{\rm Hz},T_1,  t=0) - \log\varepsilon''(f = 1{\rm Hz}, T_2,  t\rightarrow\infty)}\,.
\end{equation} 
For any normalized relaxation function $R(t)$ the Kovacs-McKenna (KM) relaxation rate $\Gamma(t)$ is defined \cite{kov63,mck94} by 

\begin{equation}\label{KM} 
\Gamma(t)\,\equiv\,-\frac{d\ln R}{dt}\,=\,-\frac{1}{R}\frac{d R}{dt}.
\end{equation}
The KM relaxation rate gives the relative change of the relaxation function with time and has the convenient property of being independent of the normalization. For a simple exponential relaxation function, $R(t)=\exp(-t/\tau)$, the KM relaxation rate is constant: $\Gamma(t)=1/\tau$. In general the KM relaxation rate changes with time. For both for temperature up and down jumps we found that $\Gamma(t)$ decreases with time (for large up jumps this does not have to be the case). A popular analytical fitting function is the stretched-exponential, $R(t)=\exp[-(t/\tau)^\beta]$ ($0<\beta<1$); this function has $\Gamma(t)=(\beta/\tau)(t/\tau)^{\beta-1}$ that decreases monotonically to zero as $t\rightarrow\infty$.

\begin{figure}
\includegraphics[width=8cm]{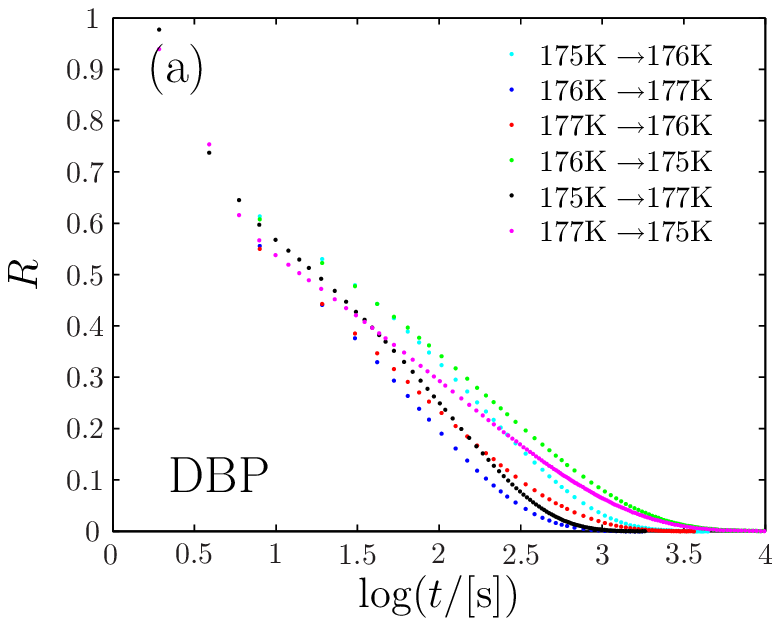}
\includegraphics[width=8cm]{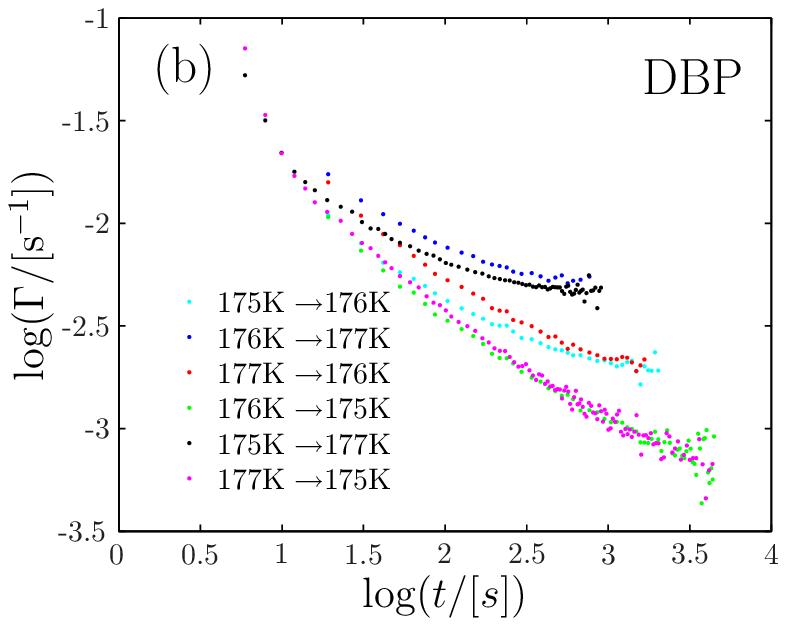}
\includegraphics[width=8cm]{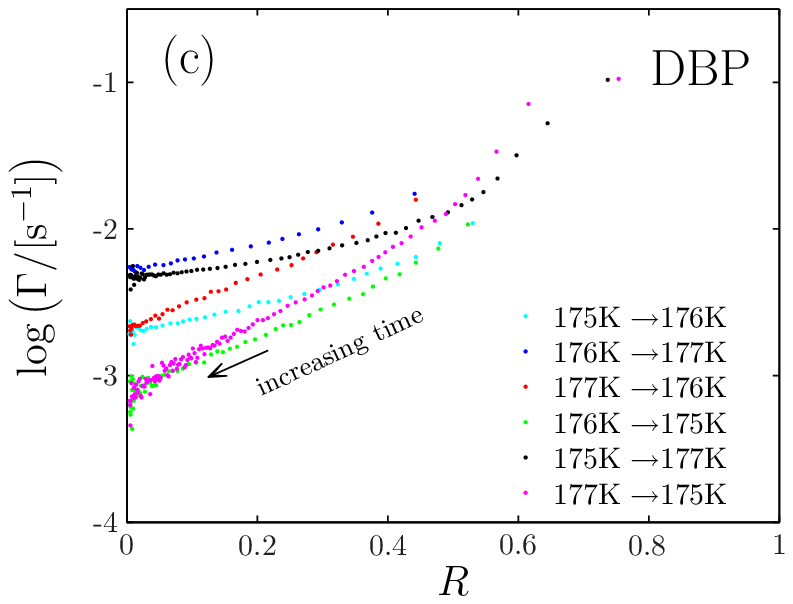}
\caption{\label{kovacsrate}
(a) Normalized relaxation functions for DBP as functions of log(time). 
(b) The Kovacs-McKenna (KM) relaxation rates $\Gamma$ defined in Eq. (\ref{KM}) for these data, as functions of log(time). Up and down jumps ending at 176 K give the same relaxation rate at long times, showing that there is no so-called expansion gap as proposed by Kovacs in 1963 \cite{kov63}. 
(c)  Parameterized plot of $(R(t),\Gamma(t))$. Again, it is seen that different temperature jumps to the same temperature approach the same relaxation rate at long times (small $R$).}
\end{figure}

Taking now DBP as an example, Fig. \ref{kovacsrate}(a) shows as functions of time the normalized relaxation functions for all six temperature jumps of Fig. \ref{raw}(a) and (b). Figure \ref{kovacsrate}(b) shows the corresponding KM relaxation rates. At long times there is considerable noise in the KM rates because the relaxation rate is difficult to determine by numerical differentiation when the noise becomes comparable to $R(t)$ \cite{str97}. In order to eliminate unreliable long-time $\Gamma(t)$ data points we introduced a cut-off at 0.5\% from equilibrium for all data sets. Despite the long-time noise it is clear that for up and down jumps ending at the same temperature ($175$, $176$, or $177$ K) the KM relaxation rates eventually approach the same number. This shows that there is no so-called expansion gap as Kovacs proposed in 1963 based on experiments monitoring relaxation by measuring volume changes \cite{kov63}. Figure \ref{kovacsrate}(c) is a parameterized plot of $(R(t),\log(\Gamma(t)))$ which -- except for the normalization of $R$ introduced here -- was the data representation originally used by Kovacs \cite{kov63}. Again, it is clear that up and down jumps to the same temperature approach the same KM relaxation rate at long times ($R\rightarrow 0$). The existence of an expansion gap has been a matter of debate \cite{ren87,str97,mck95,mck99}. Kolla and Simon recently concluded, however, that there is no expansion gap for $t\rightarrow\infty$; they attributed the reported expansion gap to the fact that Kovacs' was unable to examine departures from equilibrium that were small enough to show the convergence of time scales \cite{kol05}.

\section{The internal clock hypothesis}\label{internal}

For interpreting the data we use the Tool-Narayanaswamy (TN) formalism, which dates back to Tool's works in the 1940's and matured with Narayanaswamy's seminal paper from 1971 \cite{too46,hop58,nar71,sch86}. The TN formalism interprets aging in terms of a so-called material time. The main feature of the TN formalism is that it describes aging in terms of a linear convolution integral, even when the aging is highly nonlinear. The formalism generally works well, although from a fundamental point of view it is still somewhat of a mystery why this is.

Lunkenheimer and coworkers recently studied aging also by monitoring the dielectric loss \cite{lun05,lun06,weh07}. They found that the relaxation curves $R(t)$ to a good approximation may be described by a stretched-exponential relaxation function which, as a new feature, introduces a time-dependent characteristic time $\tau(t)$: $R(t) =  \exp \left[-(t/\tau(t))^\beta\right]$. The nonlinear stretching exponent $\beta$ was found to be identical to that derived from the linear dielectric relaxation function. This is a novel approach to aging studies. However, it does not lend any obvious physical interpretation to $\tau(t)$, which has the appearance of a ``global'' averaged relaxation time representing the entire aging process until time $t$. 

The TN approach's material time may be thought of as time measured on a clock that changes its rate as sample properties evolve with time. The material time is analogous to the proper time concept of relativity theory, the reading on a clock following a (possibly accelerated) observer's world line; from an inertial system one would say that the clock rate varies with the observer's velocity, but the moving observer would dispute this. The  the existence of a material time is an old idea that predates Narayanaswamy; thus the well-known time-temperature superposition concept may be regarded as a ``linear'' internal clock hypothesis. Narayanaswamy's brilliant insight was to generalize this to describe aging, which is a  highly nonlinear phenomenon.

The material time, which will be denoted by $\til$, is time measured using a time unit that itself evolves with time. If the structural clock rate is denoted by $\gamma_s(t)$, the material time $\til$ is defined \cite{too46,ren87,str97,mck99,kol05,hop58,mck95,nar71,sch86} by

\begin{equation}\label{mat_time_def} 
d\til\,=\,\gamma_s(t) dt\,.
\end{equation}
This means that

\begin{equation}\label{mat_time_def_int} 
\til(t)\,=\,\int_{-\infty}^t\gamma_s(t) dt\,,
\end{equation}
where the lower bound is, of course, arbitrary. The TN formalism is standard for interpreting aging experiments and used routinely used in industry for predicting aging effects \cite{tri86,hod97,hor10}. Nevertheless, it is not known whether -- and in which sense -- the internal clock exists, or if it should merely be regarded as a convenient mathematical construction. 

According to the TN formalism, for all temperature jumps applied to a given system -- small or large, up or down -- the normalized relaxation function is a unique function of the material time that has passed since the jump was initiated at $\til=0$: $R=R(\til)$. In applications of the TN formalism one often allows for different material times to control the aging of different quantities (with the function $R(\til)$ varying with the quantity that is being monitored). But if an internal clock {\it really} exists, a common  material time must control all relaxations. In particular, the relaxation of the clock-rate activation energy itself during aging must be controlled by the same material time that controls the dielectric aging process (details are given below). A major point of this paper is to check against experiments the consequences of assuming that an internal clock exists. The next subsections develop a theory for testing this.

Determining the structural clock rate $\gamma_s(t)$ in the TN formalims usually involves some mathematical modelling, fitting of data, or assumption regarding what controls the relaxation \cite{moy76,sch86,nar71,ols98}. In Sec. \ref{howto} we develop a test of the internal clock hypothesis which does not require such procedures, but proceeds directly from data without explicitly determining $\til(t)$. First, however, it is necessary to define precisely both the dielectric relaxation rate in an out-of-equilibrium situation and the structural relaxation clock rate.

\begin{figure}
\includegraphics[width=8cm]{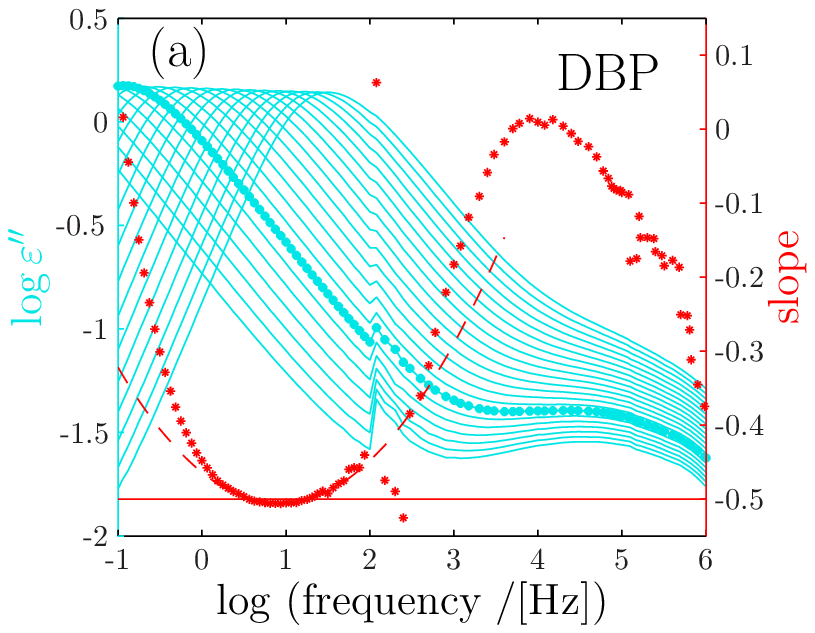}
\includegraphics[width=7.6cm]{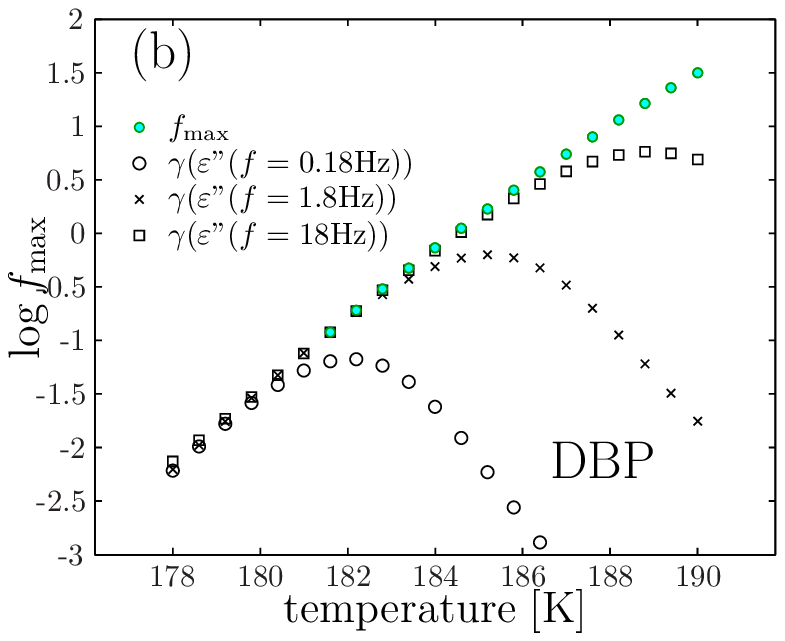}
\caption{\label{determineCR}
 (a) Illustration of the procedure used to determine the inverse power-law exponent $\beta$, which is identified as the minimum slope of the dielectric loss curve in a log-log plot at the temperature where the loss-peak frequency is 0.1 Hz (blue dotted curve). The red data points give the numerical slopes of this curve, and the red dashed curve is a parabola fitted to the bottom points of the slope; the analytic minimum of the parabola determines the minimum slope \cite{nie09}.
(b) The loss-peak frequencies determined from the equilibrium spectra (green) and the predicted peak-positions using Eq. (\ref{structural}) below (corresponding to $\gamma_d$) at different  measuring frequency. The curves line up at low temperatures, showing that this procedure determines the correct loss-peak frequency.}
\end{figure}

\begin{table}[h!] 
\begin{tabular}{l|ccccc}
&	DBP &	DEP &	2,3-epoxy &	5-PPE &	TPP \\
\hline	 	 	 	 	 
$\beta$ &	0.506 &	0.483 &	0.550 &	0.507 &	0.495

\end{tabular}
\caption{\label{exponents}The high-frequency slopes $\beta$ used in the data analysis.}
\end{table}

\subsection {Defining the dielectric relaxation rate for out-of-equilibrium situations}

The five liquids studied are all good glass formers that obey time-temperature superposition (TTS) for their main (alpha) process to a good approximation. Moreover, they all have a high-frequency decay of the loss that to a good approximation may be described by a power-law, $\varepsilon''(f)\propto f^{-\beta}$, where $\beta$ is close to $1/2$. It was conjectured some time ago that a high-frequency exponent of $-1/2$ reflects the generic properties of the pure alpha process obeying TTS (i.e., whenever the influence of additional relaxation processes is negligible) \cite{ols01}, a conjecture that was strengthened by a recent study involving more than 300 dielectric spectra \cite{nie09}. For the below data analysis the exponent $\beta$ was identified as the minimum slope \cite{nie09} of the log-log plotted dielectric loss curve above the loss peak, evaluated at the temperature where the loss peak is 0.1 Hz (Fig. \ref{determineCR}). The $\beta$ values thus obtained are listed in Table \ref{exponents}.

The inverse power-law high-frequency dielectric loss, shown for DBP in Fig. \ref{ttsdbp}(a), is used to monitor the dielectric relaxation rate $\gamma_d(t)$ as the structure ages following a temperature jumps. This is done by proceeding as follows. First, we define the dielectric relaxation rate for the equilibrium liquid, $\gamma_d$, as the angular dielectric loss-peak frequency:

\begin{figure}
\includegraphics[width=8cm]{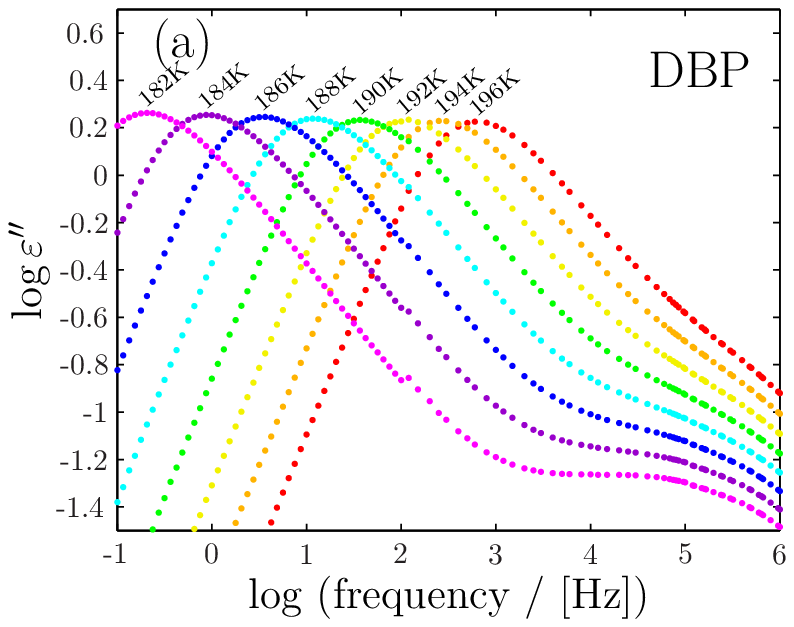}
\includegraphics[width=8cm]{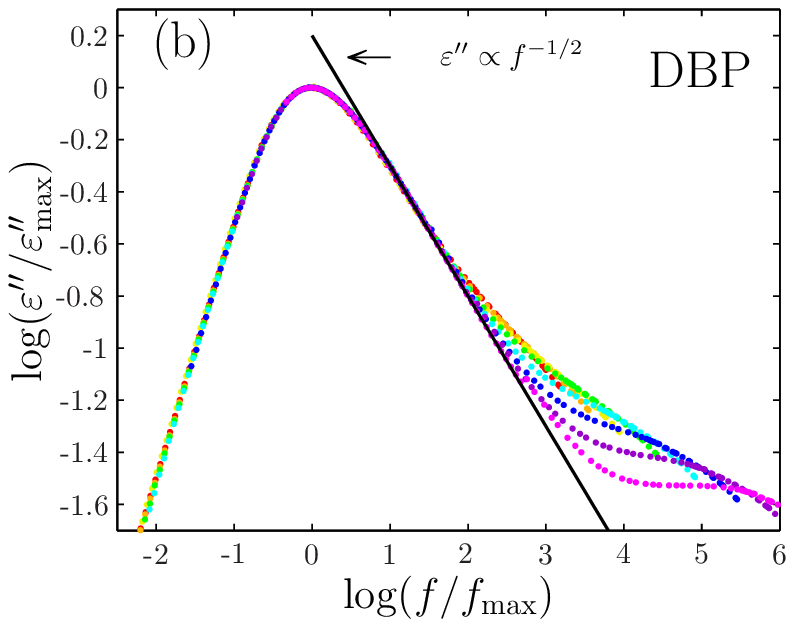}
\includegraphics[width=8cm]{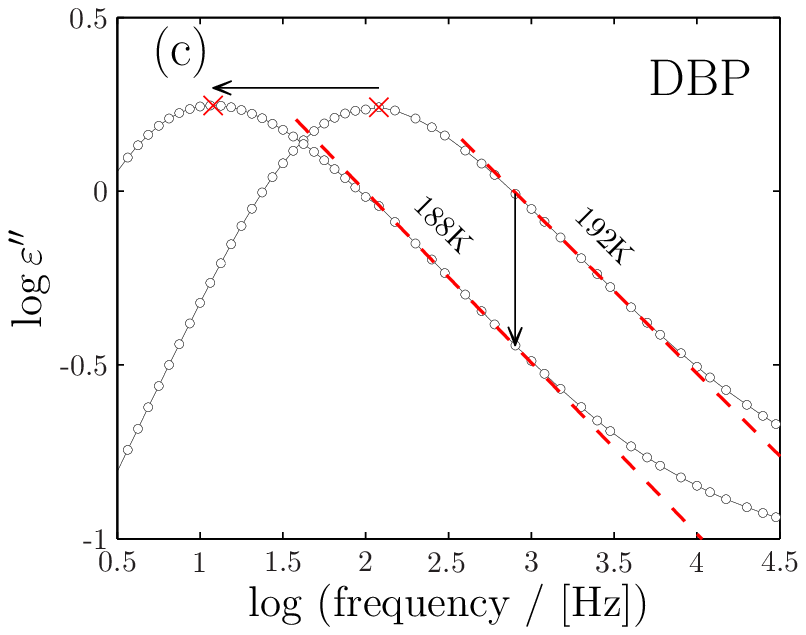}
\caption{\label{ttsdbp}
(a) Dielectric loss spectra for DBP above $T_g$ (i.e., equilibrium data). 
(b) TTS plot of the same spectra illustrating that the high-frequency wing of the alpha (main) process approaches a slope of $-1/2$ as temperature is lowered \cite{ols01,nie09}. All five liquids have high-frequency slopes close to $-1/2$, but this fact is not important for the analysis.
(c) Illustration of how one utilizes the fact that the loss varies as $f^{-\beta}$ at high frequencies to measure the dielectric loss-peak frequency that by definition gives the dielectric clock rate, also during aging (Eq. (\ref{structural})).}
\end{figure}

\begin{equation}\label{gd_def}
\gamma_d\equiv 2\pi f_m\,,
\end{equation}
where $f_m$ is the loss-peak frequency. If temperature is lowered in a step experiment, the dielectric loss curve gradually moves to lower frequencies as the system ages and relaxes to equilibrium. How to define a dielectric relaxation rate $\gamma_d$ for this out-of-equilibrium situation? It is not possible to continuously monitor the entire loss curve, because the aging takes place on the same time scale as that of the dielectric loss, implying that linear-response measurements around the loss peak frequency are not well defined (i.e., a harmonic input does not result in a harmonic output). To circumvent this problem, the intuitive idea is that how much the dielectric relaxation rate has changed may be determined from how much the loss has changed at some fixed frequency in the high-frequency power-law region (Fig. \ref{ttsdbp}(c)). Mathematically, this corresponds to defining $\gamma_d(t)$ from the high-frequency equilibrium expression as follows

\begin{equation}\label{clockrate}
\varepsilon''(f,t)\,\propto\, (f/\gamma_d(t))^{-\beta}\,.
\end{equation}
Thus by probing the dielectric loss at the fixed frequency $f$, the dielectric relaxation rate may be determined during aging from

\begin{equation}\label{structural}
\log\gamma_d(t)\,=\,\frac{1}{\beta}\log \varepsilon''(f,t) +A\,.
\end{equation}
The calibration constant $A$ is found by using equilibrium data from higher temperatures where the loss peak is within the observable frequency range. It is not a new idea to monitor aging by measurements at a frequency much larger than the reciprocal structural relaxation time; for instance Struik long ago discussed the proper protocols for doing this \cite{str78}. 

Although the above ideas seem straightforward, from a conceptual point of view one may question the validity of the concept of a dielectric relaxation rate in a situation where the structure ages on the same time scale as the dipoles relax. In order to specify the precise assumptions needed to  justify defining $\gamma_d(t)$ via Eq. (\ref{structural}), we reason as follows. According to linear-response theory, for a system in thermal equilibrium the measured output is calculated from a convolution integral involving the input before the measuring time. A convenient way to summarize time-temperature superposition (TTS) for the equilibrium liquid is to formulate the convolution integral in terms of a dielectric ``material'' time $\til$: If $\gamma_d$ is the equilibrium liquid's dielectric relaxation rate (Eq. (\ref{gd_def})), the dielectric material time is defined from the actual time $t$ by  

\begin{equation}\label{t_til_tts}
\til\,=\, \gamma_d\, t\,.
\end{equation}
In terms of $\til$, since in a standard dielectric experiment the input variable is the electric field $\bf E$ and the output is the displacement vector $\bf D$, the convolution integral is of the form

\begin{equation}\label{2}
{\bf D}(\til)\,=\,\int_0^\infty {\bf E}(\til-\til') \psi(\til')d\til'\,.
\end{equation}
Equation (\ref{2}) describes TTS because it implies that, except for an overall time/frequency scaling, the same frequency-dependent dielectric constant is observed at different temperatures (we ignore the temperature dependence of the overall loss, an approximation which introduces a relative error into the data treatment well below 1\% over the range of temperatures studied).  

In Eq. (\ref{2}), which applies at equilibrium whenever TTS applies, the dielectric material time is defined from the actual time by scaling with $\gamma_d$ (Eq. (\ref{t_til_tts})). In the out-of-equilibrium situation following a temperature jump, the simplest assumption is that Eq. (\ref{2}) also applies, however with a generalized dielectric material time involving a time-dependent dielectric relaxation rate $\gamma_d(t)$, i.e.,

\begin{equation}\label{3}
d\til\,=\,\gamma_d(t) dt\,.
\end{equation}
As the system gradually equilibrates at the new temperature, the dielectric relaxation rate $\gamma_d(t)$ approaches the equilibrium liquid's loss-peak angular frequency at the new temperature. The equilibrium liquid's power-law dielectric loss $\varepsilon''\propto f^{-\beta}$ applies in a range of frequencies obeying $f \gg f_m$. Since by Eq. (\ref{2}) $\varepsilon(\tom)=\int_0^\infty\psi(\til')\exp(-i\tom\til')d\til'$ where $\tom=\omega/\gamma_d$, the equilibrium liquid's loss obeys $\varepsilon''\propto \tom^{-\beta}$ for $\tom\gg 1$. By the mathematical Tauberian theorem this implies that $\psi(\til')\propto(\til')^{\beta-1}$ whenever $\til'\ll 1$. The proposed generalization of Eq. (\ref{2}) to out-of-equilibrium situations now mathematically implies that the dielectric relaxation rate $\gamma_d(t)$ is given by Eq. \refeq{clockrate}. In summary, assuming the simplest generalization of TTS to out-of-equilibrium situations, a generalized dielectric relaxation rate has been defined; moreover we have shown how to measure it by monitoring the high-frequency dielectric loss at a fixed frequency using the inverse power-law approximation. 

The idea of determining an out-of-equilibrium relaxation rate directly from experimental data instead of via modelling is mathematically equivalent to the so-called time-aging time superposition \cite{leh98,bra97,oco97,lee98,oco99}. This was traditionally \cite{kov63b,mck95,bec97,ech03} implemented by first using the short-time response of, for instance, a mechanical perturbation to take a ``snap-shot'' of the structure during a volume-recovery experiment. These curves are then shifted horizontally on the time axes in order to determine the aging-time shift factors, $a_{T_f}$. Assuming time-aging time superposition, the shift factors are proportional to the structural relaxation time. Thus, the reduced time is found via an equation equivalent to Eq. (\ref{3}), $\til = \int_0^t (a_{T_f}(t'))^{-1} \,dt$ \cite{kov78}.

In the following we relate $\gamma_d(t)$ to the TN structural relaxation clock rate $\gamma_s(t)$, but first the latter quantity needs to be precisely defined.

\subsection{Defining the structural relaxation clock rate}

The structural relaxation clock rate $\gamma_s(t)$ determines the structural relaxation's material time in the TN formalism. Just as was the case for the generalized dielectric relaxation rate, it is not {\it a priori} obvious that any $\gamma_s(t)$ may be defined; the eventual test of the existence of $\gamma_s(t)$ is whether a consistent description is arrived at by assuming its existence. Assuming for the moment that this is the case, we define the structural relaxation clock rate's time-dependent activation (free) energy $E(t)$ by writing 

\begin{equation}\label{4}
\gamma_s(t)\,=\,\gamma_0 e^{-E(t)/k_BT}\,\,(\gamma_0=10^{14}{\rm s^{-1}})\,.
\end{equation}
The activation energy $E(t)$ depends on the structure and evolves during the structural relaxation. Consider the case of structural relaxation induced by a general temperature variation. According to the TN formalism the aging of the activation energy  is described by a  linear convolution integral over the temperature history involving a material time $\til_s$ defined by the analogue of Eq. (\ref{3}), 

\begin{equation}\label{mt_s_def}
d\til_s\,=\,\gamma_s(t) dt\,.
\end{equation}
Including for convenience the inverse temperature in the below equation, the linear convolution integral for the activation energy's evolution induced by a temperature variation,  $T(t)=T_0+\Delta T(t)$, is given by an expression of the form

\begin{equation}\label{5}
\Delta (E/k_BT)(\til_s)\,=\,\int_0^\infty {\Delta T}(\til_s-\til_s') \phi(\til_s')d\til_s'\,. 
\end{equation}

\subsection{Assuming the existence of an internal clock}

A main purpose of this paper is to investigate the consequences of assuming that an internal clock exists. This assumption implies that the same material time controls dielectric aging via Eq. (\ref{2}) and aging of the structural relaxation clock rate via Eq. (\ref{5}), i.e., that for any aging experiment one has

\begin{equation}\label{7}
\gamma_s(t)\,\propto\,  \gamma_d(t) \,.
\end{equation}

A clock works by counting repeated physical processes, and two clocks measure the same physical time if the number of ticks counted by the clocks are proportional for all time intervals. Thus both above-defined clock rates $\gamma_d$ and $\gamma_s$ are defined only up to a proportionality: The physical content of Eqs. (\ref{2}) and (\ref{5}) is invariant if the reduced times are redefined by multiplying by some number. Nevertheless, Eq. (\ref{7}) is not trivial; thus Eqs. (\ref{2}) and (\ref{5}) may both apply with different definitions of the reduced time. As mentioned, the TN formalism is often used assuming that different physical quantities (e.g., volume and enthalpy) relax with rates that are not proportional \cite{sch86}. 

If Eq. \refeq{7} applies, we find via Eqs. (\ref{clockrate}) and (\ref{4}) that after a temperature jump to temperature $T$ the logarithm of the measured loss is given by

\begin{equation}\label{8}
\ln\varepsilon''(f,t)\,=\,-\beta\,\frac{E(t)}{k_BT}+C \,,
\end{equation}
and that this quantity relaxes following a material time whose rate may be determined from Eq. \refeq{structural}. We proceed to derive a test of this prediction. If this is fulfilled, the internal clock hypothesis will be regarded as confirmed.

\section{A test for the existence of an internal clock}\label{howto}

In this section we show that the existence of an internal clock, i.e., the assumption that the dielectric clock rate is proportional to the structural relaxation clock rate (Eq. (\ref{7})), can be tested without evaluating $\til$ explicitly and without fitting data to analytical functions. 

First, we define a dimensionless KM relaxation rate by replacing time in Eq. (\ref{KM}) by the reduced structural relaxation time,

\begin{equation}\label{gammadim_def}
\tilde\Gamma\equiv -\frac{d\ln R}{d \til_s}\,.
\end{equation}
According to the TN formalism, for all temperature jumps $R(\til_s)$ is the same function of $\til_s$. This implies that $\tilde\Gamma(\til_s)$ is the same for all jumps. By eliminating $\til_s$, $\tilde\Gamma$ is a unique function of $R$:

\begin{equation}\label{prediction}
\tilde\Gamma\,=\,\Phi(R)\,.
\end{equation}
Thus one way of testing whether the TN formalism applies is to check whether $\tilde\Gamma$ is a unique function of the normalized relaxation function for different temperature jumps. To do this we express the dimensionless KM relaxation rate in terms of the real unit KM relaxation rate, 

\begin{equation}\label{gammatilde}
\tilde\Gamma (\til)= -\frac{d\ln   R}{dt}\frac{dt}{d\til_s} = \frac{\Gamma(t)}{\gamma_s(t)}\,.
\end{equation}
If an internal clock exists,  $\gamma_s(t)$ may be evaluated from its proportionality to the dielectric relaxation rate Eq. (\ref{7}), which is  accessible via Eq. (\ref{structural}). Note that the unknown proportionality constant in Eq. (\ref{7}) is irrelevant because, as mentioned, clock rates are only defined up to a proportionality constant (in Sec. \ref{calibrating} we discuss the possibility of absolute calibration of the structural and dielectric clock rates). In summary, if $\gamma_s(t)\propto\gamma_d(t)$, via Eq. (\ref{gammatilde}) $\tilde\Gamma$ may be calculated directly from a temperature jump experiment's data, since $\Gamma(t)$ and $\gamma_d(t)$ are determined both from $\ln\varepsilon''(f,t)$  via Eqs. (\ref{KM}) and (\ref{structural}), respectively.

\begin{figure}
\includegraphics[width=8cm]{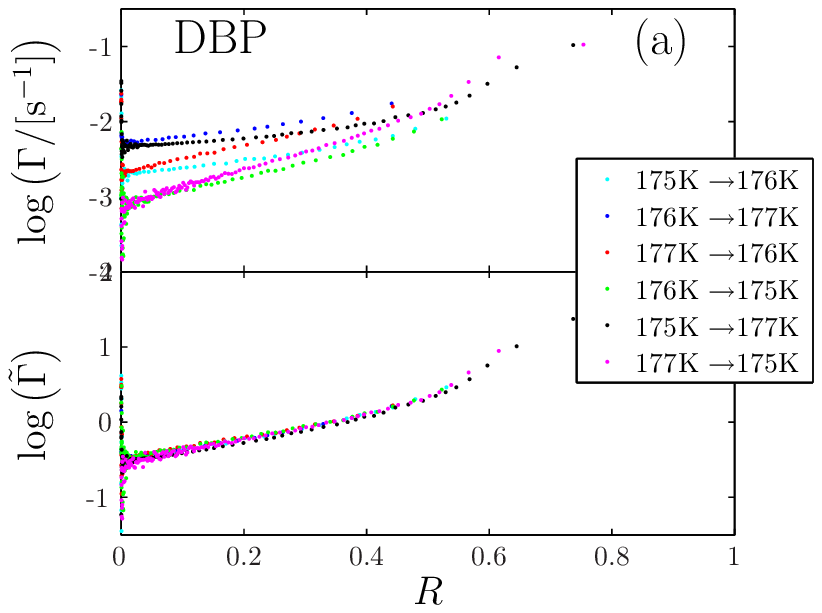}  
\includegraphics[width=8cm]{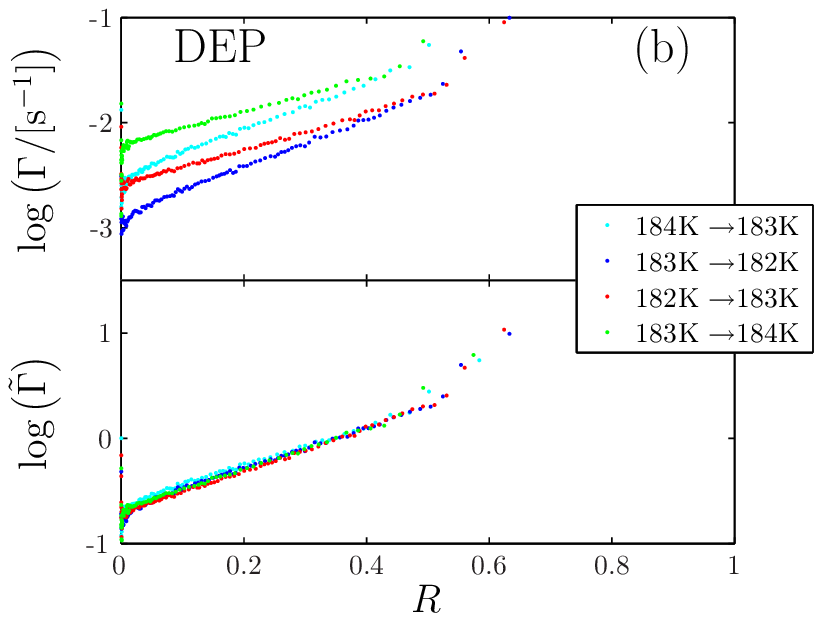}  
\includegraphics[width=8cm]{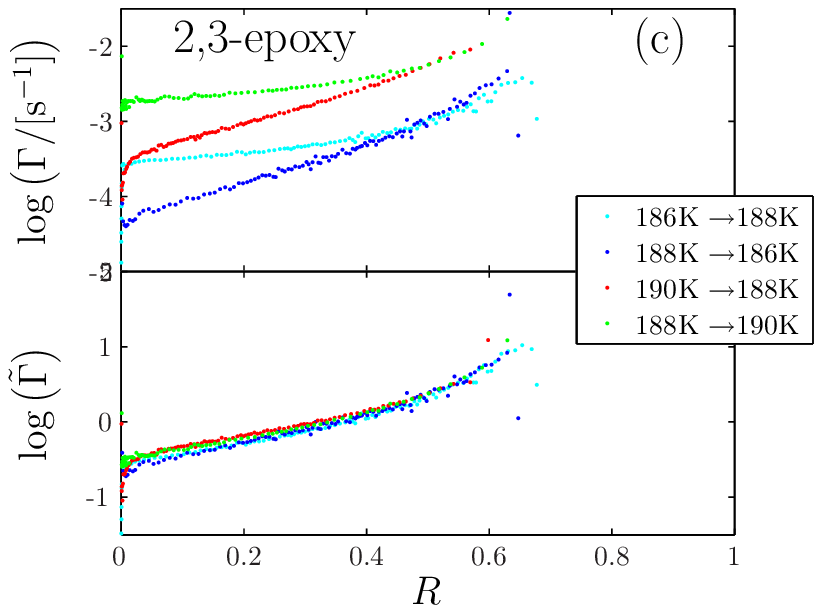} 
\includegraphics[width=8cm]{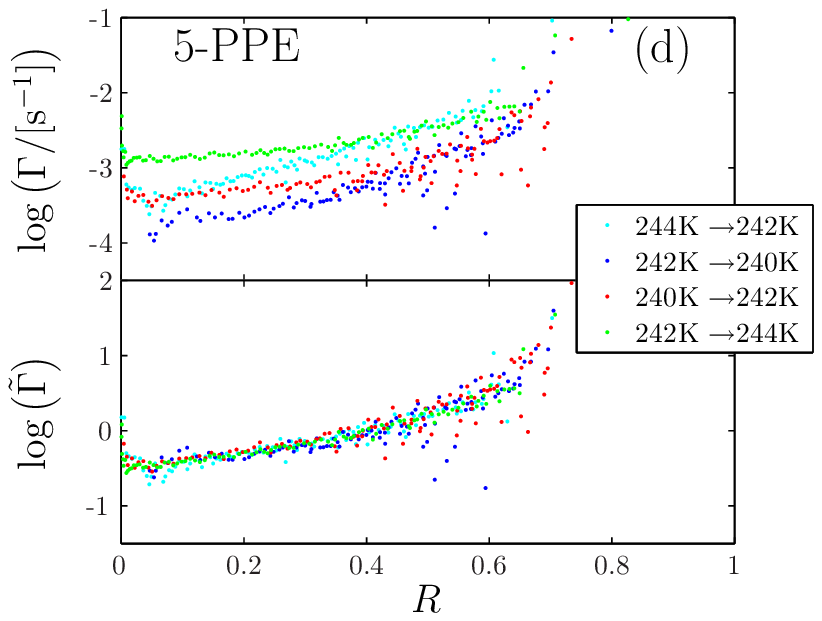} 
\includegraphics[width=8cm]{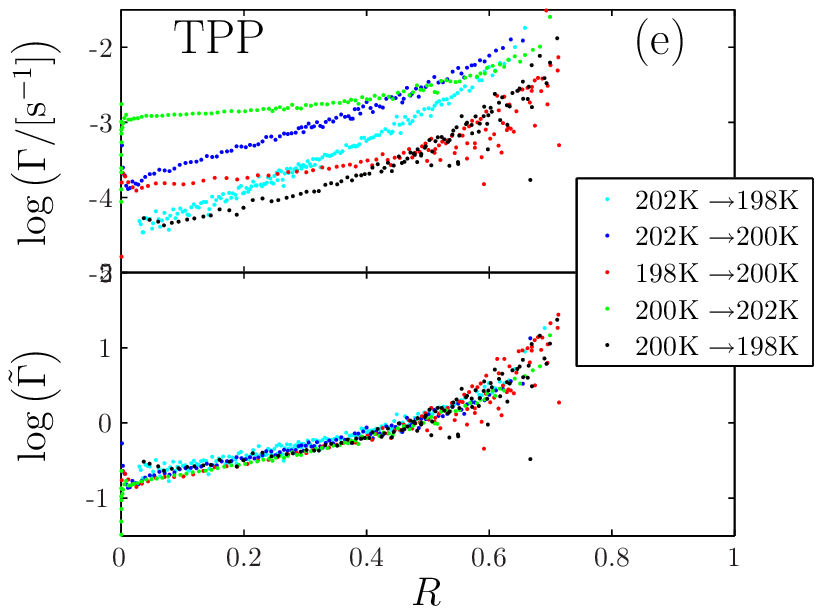}  
\caption{\label{kovacsplots}Kovacs-McKenna (KM) relaxation rates $\Gamma$ and its dimensionless version $\tilde\Gamma(\til)=\Gamma(t)/\gamma_d(t)$ (defined in Eq. (\ref{gammadim_def}) and calculated from data via the internal-clock hypothesis, $\gamma_s(t)=\gamma_d(t)$) as functions of the normalized relaxation functions $R$ for the five liquids. For each liquid the upper subfigure shows $\Gamma(R)$, the lower subfigure shows  $\tilde\Gamma(R)$. In all cases there is data collapse of $\tilde\Gamma(R)$ within experimental errors. This confirms the existence of an internal clock for these liquids.}
\end{figure}

\begin{figure}
\includegraphics[width=8cm]{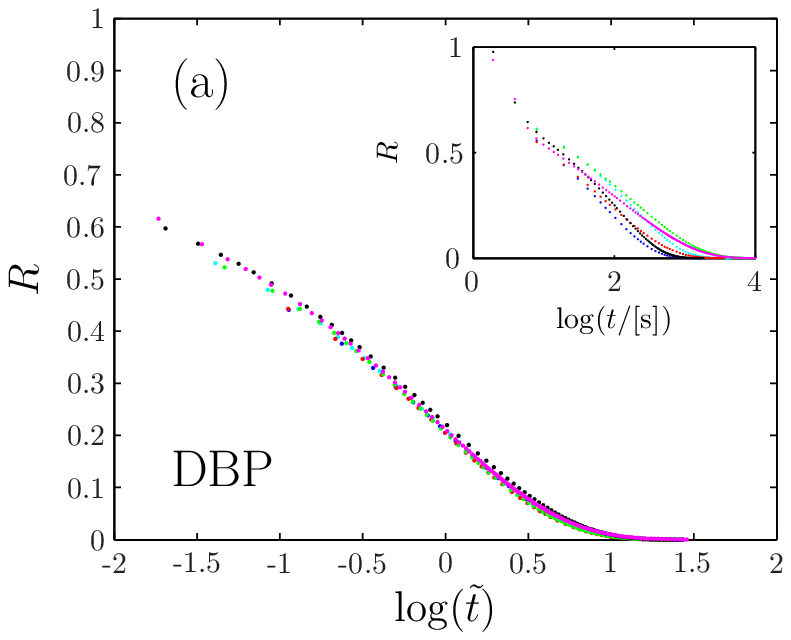} 
\includegraphics[width=8cm]{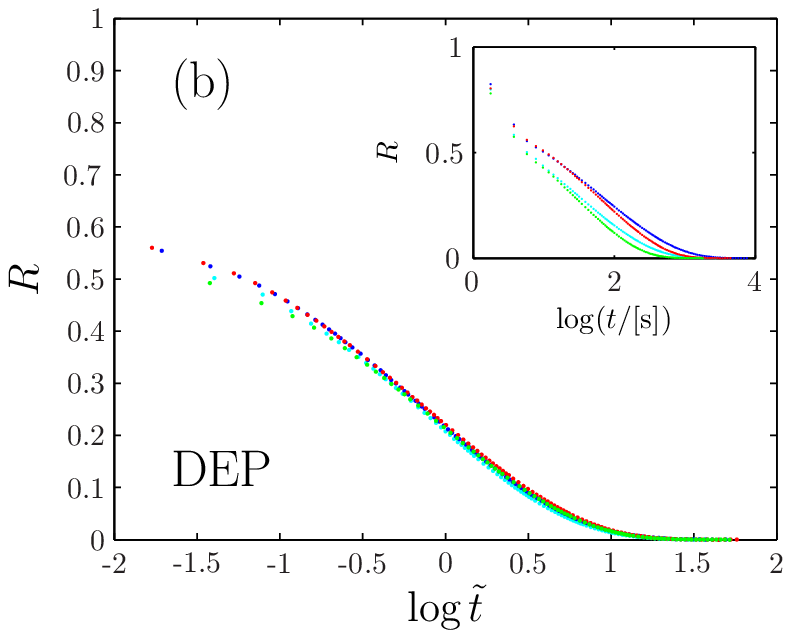}
\includegraphics[width=8cm]{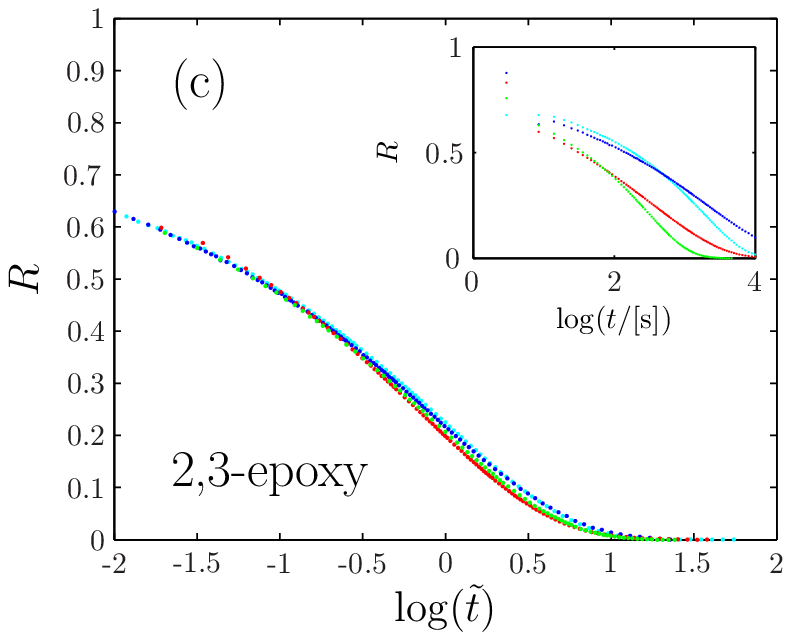} 
\includegraphics[width=8cm]{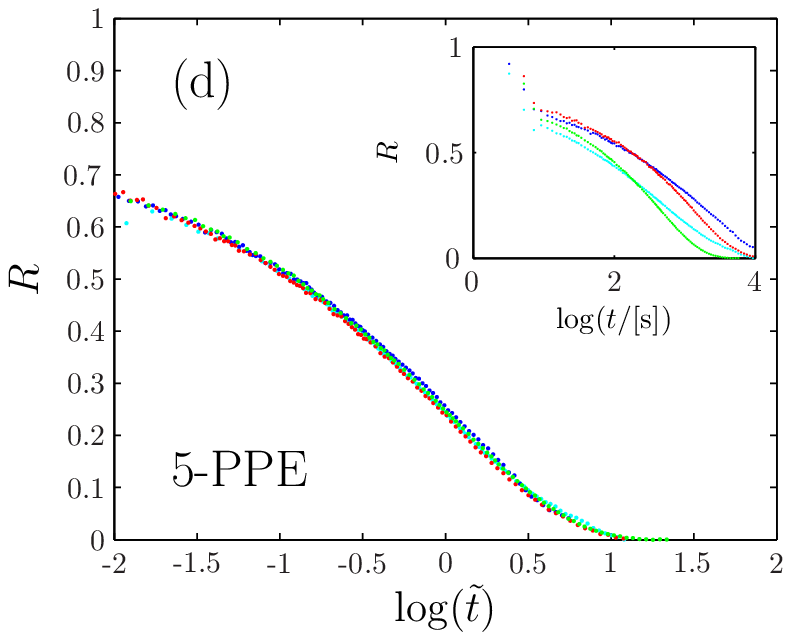} 
\includegraphics[width=8cm]{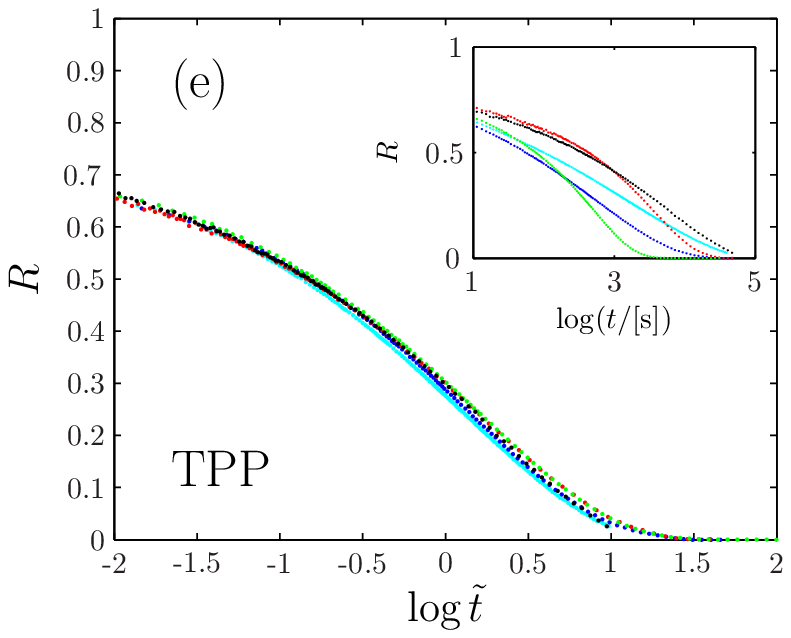} 
\caption{\label{tildeR} The traditional way of demonstrating TN data collapse by plotting the normalized relaxation functions as function of the reduced time, $R(\tilde{t})$. The insets show the normalized relaxation functions plotted against real time, $R(t)$.}
\end{figure}

Defining the proportionality constant between the two rates to be unity, $\gamma_s(t)=\gamma_d(t)$, the results for the KM relaxation rates $\Gamma(R)$ and the dimensionless KM relaxation rates $\tilde\Gamma(R)$ are plotted in Fig. \ref{kovacsplots}. For all five liquids the results are consistent with the internal clock hypothesis. Even the 4K down jump for TPP -- corresponding to a clock-rate variation of almost two orders of magnitude -- falls nicely onto the master curve. The spread in KM relaxation rates as $R\cong 0$ is approached at long times reflects the already mentioned fact that relaxation rates cannot be determined reliably by numerical differentiation when the noise becomes comparable to the distance to equilibrium.  

Once the existence of an internal clock has been demonstrated, it is natural to evaluate the reduced time $\til$ explicitly by integration in order to determine $R(\til)$. As shown in Fig. \ref{tildeR} this gives the data collapse predicted by the TN formalism. For the numerical integration one must either include short-time transient points, where the sample still undergoes temperature equilibration, or omit the initial measurements. The error introduced from this uncertainty influences all values of $\til$. This is one reason to prefer the ``direct'' test of the internal clock hypothesis of Fig. \ref{kovacsplots}; another reason is that the direct test is simpler because it avoids evaluating the material time $\til$.

\section{Long-time asymptotic behavior of the structural relaxation}\label{longtime}

\begin{figure}
\includegraphics[width=8cm]{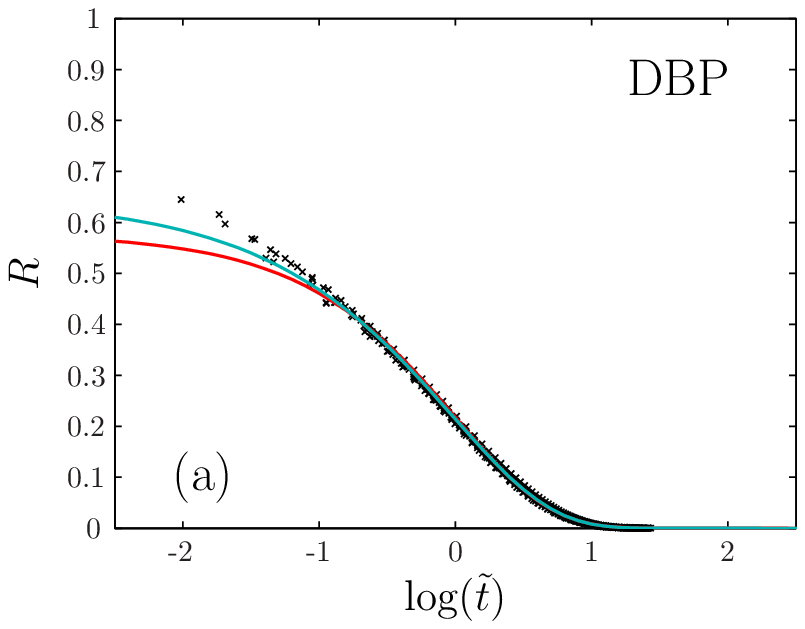}
\includegraphics[width=8cm]{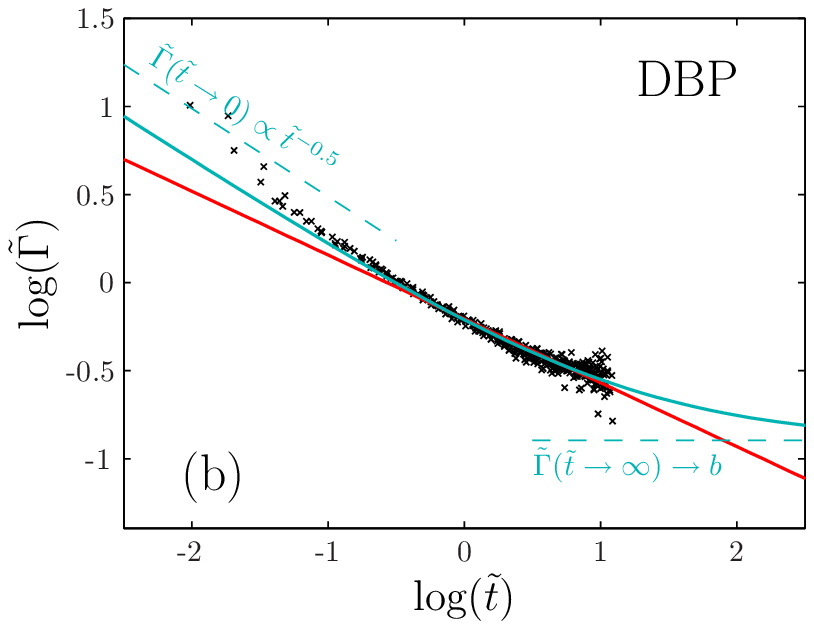}
\caption{\label{exptail}Fits of  the ``exponential $\sqrt t$ relaxation'' function (Eq. (\ref{bel2})) (blue) and the stretched exponential ($R(\til)= \exp\left(-a_{\text{se}} -(b_{\text{se}}\til)^{c_{\text{se}}}\right)$) (red) to DBP data. (a) In the standard representation showing $R$ as a function of $\log (\tilde{t})$ it is hard to distinguish the two fitting functions. (b) Relaxation rates as functions of the (reduced) times in a log-log plot. In this representation the stretched exponential function is a straight line, while the exponential $\sqrt t$ relaxation function of Eq. (\ref{bel2}) has a ``banana'' shape: at short times it gives a straight line with slope $-1/2$, at long times it bends over and eventually levels off to a constant value. The two asymptotes are marked with dashed lines. Although the measurements are noisy at long times, we conclude that the data do not follow a straight line, but have a curved shape similar to the one suggested by the exponential $\sqrt t$ relaxation function.}
\end{figure}

Inspecting the shape of the dimensionless KM relaxation rate as a function of the normalized relaxation function in Fig. \ref{kovacsplots} shows that the aging is not exponential, because that would imply a constant KM relaxation rate. The stretched-exponential function, $\exp[-\til^\beta]$, is commonly used for fitting relaxation functions. It is difficult to get reliable data on the long-time behavior of structural relaxations, but our data allow one to get such data with fair accuracy. Figure \ref{kovacsplots} shows that $\tilde\Gamma(R)\rightarrow{\rm Const.}$ at long times ($R\rightarrow 0$) for all five liquids. This is also evident from the DBP data for which Fig. \ref{exptail}(a) shows the dimensionless Kovacs plots, a stretched exponential (red line), and Eq. (\ref{bel2}) (blue line) with the values of the fit parameters listed in Table \ref{fitpar}. The  KM relaxation rate for the same data is in Fig. \ref{exptail}(b), where again is included a test of the fit by the stretched-exponential relaxation function (red straight line). Although the data become noisy at long times, they do show a bend over at long times that is inconsistent with the stretched exponential relaxation function -- the KM relaxation rates approach a finite value at long times. The blue curve in Fig. \ref{exptail} (b) is the ``exponential $\sqrt t$'' relaxation function detailed below.

\begin{table}[h!] 
\begin{tabular}{l|ccccc}
&	DBP &	DEP &	2,3-epoxy &	5-PPE &	TPP \\
\hline	 	 	 	 	 
a &	0.42 &	0.46 &	0.37 &	0.35 &	0.33 \\
b &	0.11 &	0.04 &	0.06 &	0.13 &	0.02 \\
c &	3.1 &	5.1 &	4.7 &	2.6 &	6.2 \\

\end{tabular}
\caption{\label{fitpar}Values of fitted parameters of Eq. \refeq{bel2}}
\end{table}

The fact that the KM relaxation rates converge to finite values means that the relaxation function at long times follows a simple exponential decay. To model this mathematically with as few parameters as possible, we fitted the data to the following ``exponential $\sqrt t$'' relaxation function that retains features of a stretched exponential with exponent $1/2$, but has a long-time simple exponential decay \cite{hor10,nesli}:

\begin{equation}\label{bel1}
R(\til)= \exp\left(-A - B\til -C\til^{1/2}\right)\,.
\end{equation}
Here $A$, $B$, and $C$ are fitting parameters. The number $A$ reflects the fact that, due to fast relaxations, the normalized relaxation function $R$ does not start at unity at the shortest experimentally accessible times. The case $B=0$ gives a stretched exponential with exponent $1/2$, the $C=0$ case gives an ordinary exponential decay. At short times one has $R(\til)\cong 1-A-C\til^{1/2}$, which justifies the name ``exponential $\sqrt t$ relaxation function'' (see Ref. \onlinecite{nie09} and its references to $\sqrt t$ relaxation in other contexts). Equation (\ref{bel1}) may be rewritten in the more convenient form

\begin{equation}\label{bel2}
R(\til)= \exp\left(-a - b\til -c(b\til)^{1/2}\right)
\end{equation}
where $a=A$, $b=B$, and $c=C/\sqrt{B}$. Recast in this form, it is clear that $b$ merely adjusts the time scale and that $c$ is the only genuine shape parameter.

\begin{table}[h!] 
\begin{tabular}{l|cccccc}
&	DBP &	DEP &	2,3-epoxy &	5-PPE &	TPP \\
\hline\hline	 	 	 	 	 
$\sigma_{\log(R)}^{{\rm exp}\sqrt t}$ &	0.041 &	0.029 &	0.042 &	0.023 &	0.026 \\
$\sigma_{\log(R)}^{{\rm str\, exp}}$ &	0.043 &	0.030 &	0.042 &	0.026 &	0.026 \\
\hline	 	 	 	 	 
$\sigma_{\log(\tilde{\Gamma})}^{{\rm exp}\sqrt t}$ &	0.052 &	0.028 &	0.081 &	0.172 &	0.111 \\
$\sigma_{\log(\tilde{\Gamma})}^{{\rm str\, exp}}$ &	0.074 &	0.037 &	0.092 &	0.186 &	0.116 \\

\end{tabular}
\caption{\label{sigmafit}
Test of how well the two functions fit data, where superscript ``${\rm exp}\sqrt t$'' is the exponential $\sqrt t$ relaxation function of Eq. (\ref{bel2}) and superscript ``${\rm str\, exp}$'' is the stretched exponential relaxation function. The quality of the fits is measured via the standard mean-square deviation $\sigma$ for fitting, respectively, $\log(R)$ as a function of time and $\log (\Gamma)$ as a function of time. The exponential $\sqrt t$ relaxation function provides a somewhat better fit than the stretched exponential.}
\end{table}

\section{Calibrating the dielectric clock rate}\label{calibrating}

The results obtained so far may be summarized as follows. The TN formalism predicts that the dimensionless KM relaxation rate (Eq. (\ref{gammadim_def})) is a unique function of $R$ for the relaxation towards equilibrium following any temperature jump. This can be tested only, however, if one is able to determine the structural relaxation clock rate $\gamma_s(t)$. This can be done either by some assumption about the clock rate's structure dependence -- a common procedure -- or, as above, by the internal clock hypothesis, $\gamma_s(t)\propto\gamma_d(t)$, where the dielectric relaxation rate is determined from data via Eq. (\ref{structural}). The data do collapse as predicted by the internal clock hypothesis, confirming the existence of such a clock for all five liquids. 

\begin{figure}
\includegraphics[width=8cm]{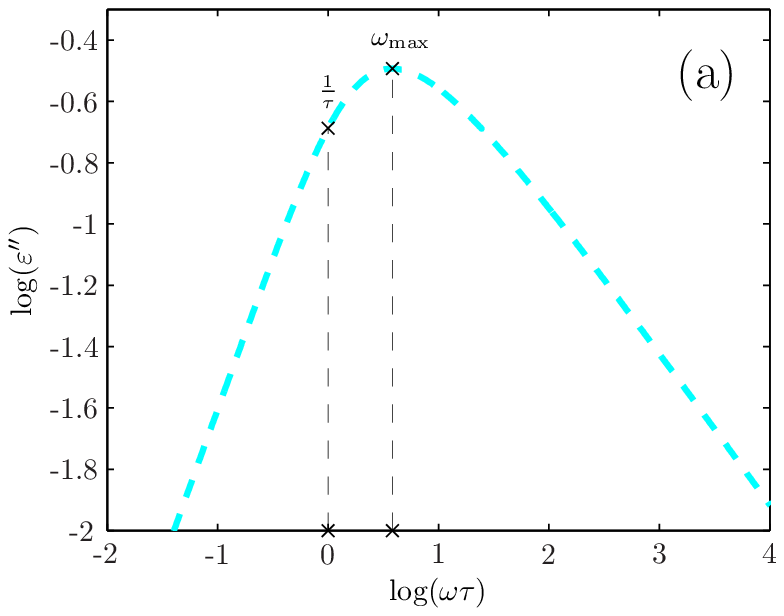}
\includegraphics[width=8cm]{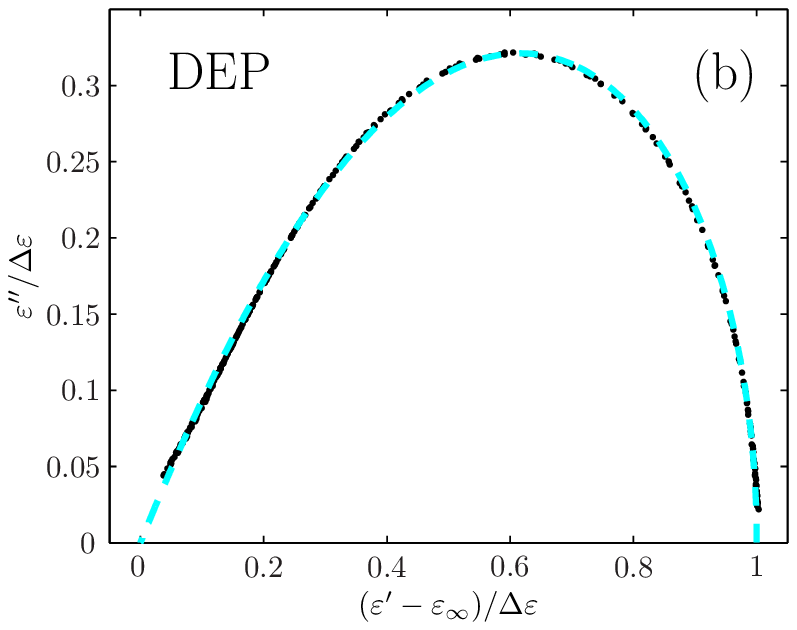}
\caption{\label{fitex}
(a) Illustration of the difference between the two calibrations of the dielectric relaxation rate $\gamma_d$, using either the loss-peak angular frequency or the rate of the long-time exponential decay of the dipole autocorrelation function giving the low-frequency Debye behavior.
(b) Normalized Cole-Cole plot of the dielectric loss of DEP (black dots) versus that of the ``exponential $\sqrt t$ relaxation'' model (Eq. (\ref{bel2})) used to fit the dielectric data at the following temperatures: 206 K, 207 K, 208 K, 209 K, 210 K, 211 K (blue dashed line).}
\end{figure}

As emphasized, a clock rate is determined only up to a proportionality constant, i.e., two clocks measure the same physical time if their numbers of ``ticks'' are proportional for all time intervals. Still, one may ask whether some sort of absolute calibration of the dielectric and structural relaxation clock rates is possible. We defined the dielectric relaxation rate in equilibrium, $\gamma_d$, as the dielectric angular loss-peak frequency (Eq. (\ref{gd_def})). This is convenient because the loss-peak frequency can easily be determined accurately. A characteristic feature of the dielectric losses of supercooled organic liquids is their pronounced asymmetry: Whereas the loss decays as a non-trivial power-law above the loss-peak frequency, at low frequencies the loss almost follows the Debye function ($\varepsilon''(\omega)\propto\omega$). Via the fluctuation-dissipation theorem the low-frequency behavior corresponds to a simple exponential long-time decay of the equilibrium dipole autocorrelation function. Inspired by the recent work of Gainaru {\it et al.} \cite{gai09} it is obvious to ask whether redefining $\gamma_d$ to be the rate of this long-time decay would imply that $\tilde\Gamma\rightarrow 1$ asymptotically at long times. In other words: Is the long-time exponential structural-relaxation clock rate equal to the exponential long-time decay of the equilibrium dipole autocorrelation function? Because the liquids studied here all obey TTS, such a recalibration of $\gamma_d$ corresponds to multiplying each liquid's equilibrium $\gamma_d$ (defined by Eq. (\ref{gd_def})) by a fixed constant. This is illustrated in Fig. \ref{fitex}.

\begin{figure}
\includegraphics[width=8cm]{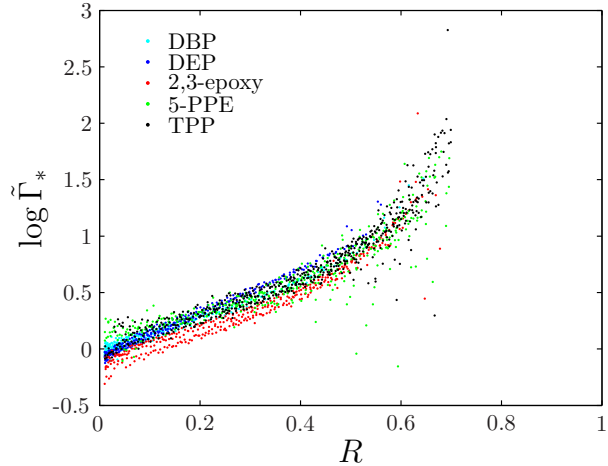}
\caption{\label{kovacscorr}Dimensionless Kovacs plots including data for all temperature jumps of the five liquids, using the alternative calibration of the dielectric relaxation rate corresponding to scaling data with the long-time dielectric relaxation rate. This procedure ``lifts'' the curves of Fig. \ref{kovacsplots} such that the dimensionless KM relaxation rates all terminate at approximately one at long times ($R\rightarrow 0$).}
\end{figure}

For each liquid the recalibration constant is obtained as follows. Assuming Eq. (\ref{bel2}) for the equilibrium dipole autocorrelation function, the liquid's dielectric loss was fitted by the Laplace transform of the negative time-derivative of this function, which interestingly provides an excellent fit to the dielectric data of all five liquids (compare Fig. \ref{fitex}(b)). In Fig. \ref{kovacscorr} we show the result of applying this recalibration of the dielectric relaxation rate in the analysis of Sec. \ref{howto}. Within experimental uncertainties all recalibrated KM relaxation rates converge to one at long times ($R\rightarrow 0$). This suggests an underlying unity in the description of aging for the liquids examined in this paper.

\begin{figure}
\includegraphics[width=9.5cm]{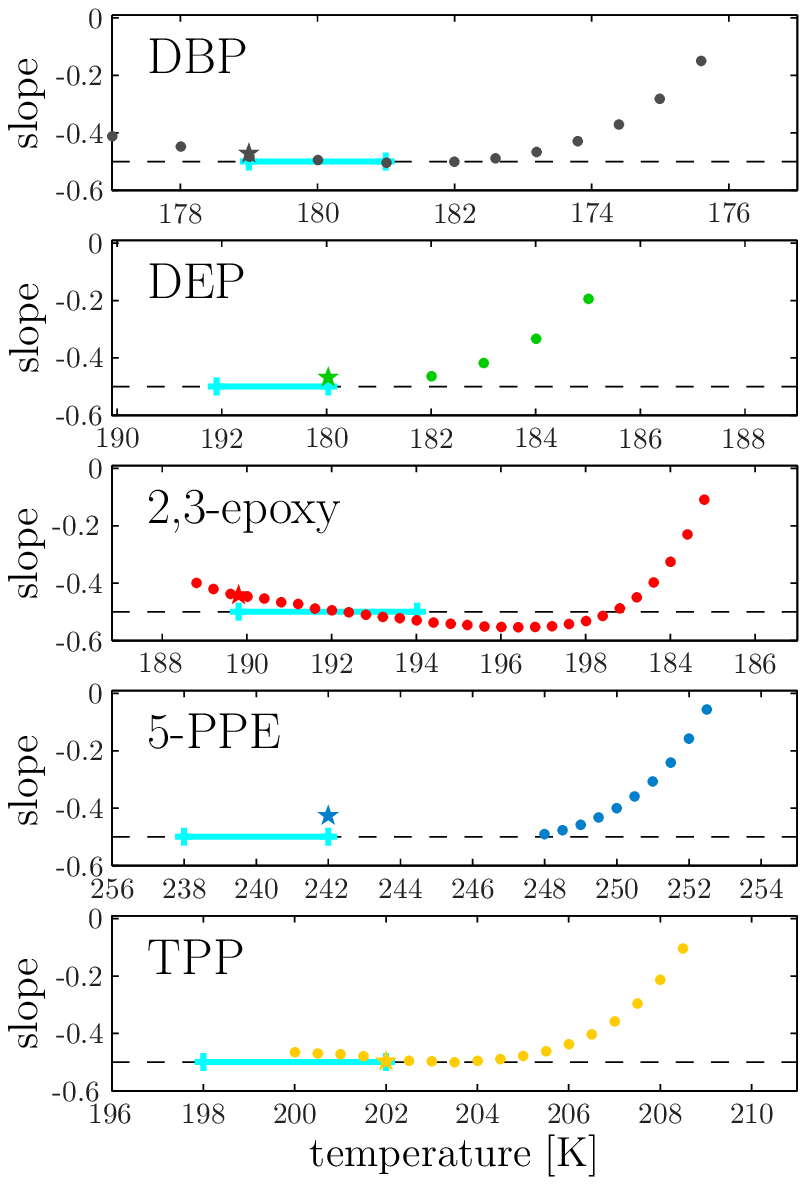}
\caption{\label{slopes} The slopes of the equilibrium log-log plotted dielectric losses at the measuring frequencies as functions of temperature. The aging interval is marked with a blue dashed line. For each frequency there is a  temperature window where the slope is constant. In this way the measuring frequency and temperature jumps can be fine tuned such that the proposed method for finding the clock rate applies.}
\end{figure}

\section{Concluding remarks}\label{sum}

We have shown how the internal clock hypothesis can be checked in a test that neither involves free parameters nor the fitting of data to some mathematical expression. The test is based on assuming the standard Tool-Narayanaswamy formalism for structural relaxation studied by monitoring the liquids' dielectric loss at a fixed frequency in the Hertz range, following temperature up and down jumps. Based on data for five organic liquids we conclude that: 1) All liquids age consistent with the TN formalism; 2) All liquids have an internal clock; 3) No liquid exhibits an expansion gap; 4) All liquids have exponential long-time relaxation; 5) The long-time structural-relaxation clock rate equals that of the long-time simple exponential decay of the dipole autocorrelation function.

Our finding that the liquids have exponential long-time relaxation is consistent with several classical viscoelastic and aging models; for instance is the famous KAHR model \cite{kov78} based on a box distribution of relaxation times, which implies the existence of a longest relaxation time and thus an exponential long-time relaxation. It is also worth emphasizing that, in contrast to reports for other materials (e.g., oxide glasses) where there is evidence that the material clock does not tick the same way for all processes, the data presented here are consistent with the existence of a unique material time. We have shown that the structural relaxation rate is proportional to the dielectric relaxation rate for five organic supercooled liquids, and the fact that the structural relaxation was monitored by measuring the dielectric loss is, in our opinion, probably not important. Nevertheless, it would be interesting to study for instance volume relaxation for the same liquids to investigate whether there really is a common material clock for these liquids. We finally note that, in contrast to the well-known TNM  formalism of Moynihan {\it et al.} \cite{moy76}, the analysis applied here does not require one or more fictive temperatures. In this sense our approach is closer in spirit to the KAHR approach (which is known, however, to be mathematically equivalent to the TNM formalism).

\begin{figure}
\includegraphics[width=8cm]{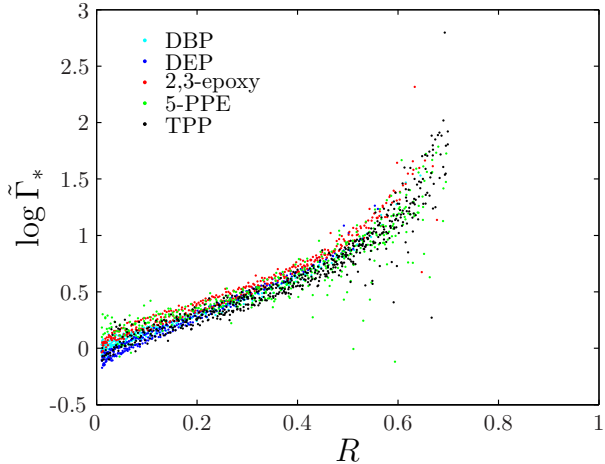}
\caption{\label{reann} Dimensionless Kovacs plots including data for all temperature jumps of the five liquids, using the alternative calibration of the dielectric relaxation rate corresponding to scaling data with the long-time dielectric relaxation rate and assuming for all liquids the high-frequency exponent of the dielectric loss $-1/2$ \cite{nie09}.}
\end{figure}

The emphasis of the data analysis was on using data directly without having to fit to analytical functions. This is why we determined the dielectric clock rate from the loss-peak angular frequency (Eq. (\ref{gd_def})) and the exponent $\beta$ as the minimum slope of the dielectric loss at the temperature where the loss peak frequency is 0.1 Hz (Table \ref{exponents}). If this purist approach is relaxed a bit, however, further interesting features appear. Thus if the dielectric clock rate is instead determined from the dielectric loss' low-frequency Debye-like behavior, all KM relaxation rates converge to unity at long times (Fig. \ref{kovacscorr}). Moreover, since the minimum slopes are not completely temperature independent, but converge to the in Ref. \onlinecite{nie09} conjectured generic value of $-1/2$ at the lowest temperatures (Fig. \ref{slopes}), one may ask what happens if the exponent $\beta$ of Eq.   (\ref{structural}) is replaced by $-1/2$. The result of repeating the entire analysis with this high-frequency exponent is shown in Fig. \ref{reann}. The main effect is to lift the 2,3-epoxy data, the liquid whose exponent $\beta$ was furthest from $-1/2$. Since the long-time structural relaxation clock rate, if identical to the redefined dielectric relaxation rate, should approach the latter from above, this figure is consistent with the conclusion that the two rates are identical.

By modern micro-engineering it should be possible to extend aging experiments to even shorter times, thus making it realistic to perform a series of ideal temperature-jump experiments over just hours. When this is eventually realized, we think it is not unlikely that aging studies could become routine on par with, e.g., present-day dielectric measurements.

\section*{{\bf APPENDIX:} $\,\,$Systematic errors and noise}\label{errors}

We discuss here the some sources of errors of the data and the analysis presented in this study. For a general and systematic analysis of errors and noise of the measurement we refer to Refs. \onlinecite{iga08a} and \onlinecite{iga08b}.

\begin{figure}
\includegraphics[width=7.5cm]{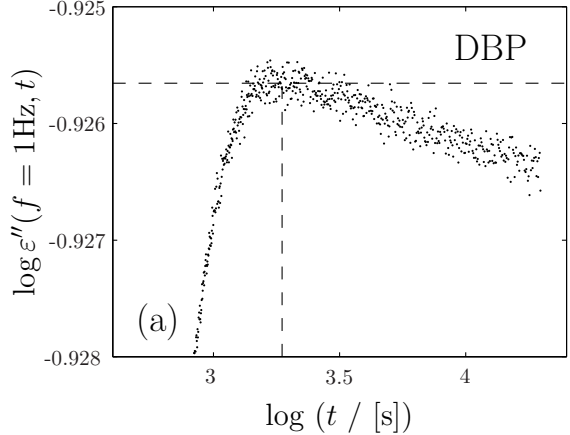}
\caption{\label{drift} Zoom of the tail of the 2K up jump to 175K for DBP (data of Fig. \ref{raw}(b)). There is a small drift at long times, which is due to the slow (compared to the alpha relaxation time) radial flow of the liquid in the measuring cell. The dashed vertical line marks where the data were cutoff and the dashed horizontal line marks the $\epsilon''(t\rightarrow\infty)$ level used in the analysis.}
\end{figure}

The geometry of the measuring cell (disc radius much larger than disc separation) introduces an extremely slow radial contraction which in equilibrium dielectric measurements can be neglected. For aging experiments it poses a problem because it introduces a small drift at long times, which distorts the curve shape of the aging relaxation function and complicates the determination of the value approached at long times. In Fig. \ref{drift} a zoom of the tail of the DBP data from Fig. \ref{raw}(b) is shown. The drift is small, but clearly visible. After a temperature step the curve should level off to a constant (equilibrium) value, instead the curve appears slightly slanted. The drift coming from the initial quench may be reduced by annealing for long time before  starting a measurement, which we did (typically over several weeks).

A further source of error is that in some cases we also observe a small overshoot when approaching equilibrium. We do not currently have an explanation for this, but it may be due to something other than the drift. Whenever a small drift or an overshoot was present, we chose to cut the data shortly after reaching its maximum/minimum and $\varepsilon''(t\rightarrow \infty)$ was adjusted accordingly. This is illustrated in Fig. \ref{drift} where the $\varepsilon''(t\rightarrow \infty)$ is marked by a horizontal dashed line and the cut-offs by a vertical dashed line. Ideally, of course, the level approached from above and below should be identical, but the deviation is in the permille range.

The signal-to-noise ratio depends on the (dielectric) relaxation strength (corresponding to the absolute level of the dielectric loss) of the liquids studied. Thus, there is more noise in the data for TPP and 5-PPE, which have relatively small dielectric relaxation strengths, compared to DBP, DEP, and 2,3-epoxy, which have larger relaxation strengths. 

\begin{figure}
\includegraphics[width=7.5cm]{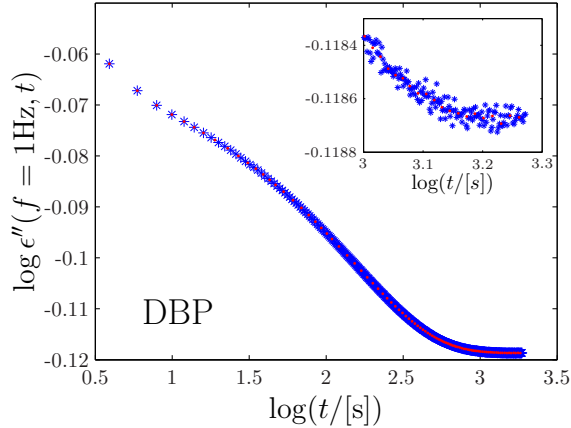}
\caption{\label{smooth}Illustration of the smoothing procedure described in the text. The algorithm minimizes the distortion of the curve shape by averaging over few (or no) data points in the beginning and more data points towards the end of the relaxation.}
\end{figure}

Although the precision of a dielectric measurement is high with barely any visible noise in the relaxation curve, we still encounter noise problems when taking the numerical derivatives of these curves. Averaging over even few data points distorts the curve shapes at short times, but it is necessary (and also less problematic) to average over more data points in the long time tails of these curves. To deal with this problem we designed an algorithm to average over a number of data points that increases with aging time, i.e., no averaging of the first data points and ending up averaging over 8 (in the case of DBP and 2,3-epoxy) or 16 (in the case of DEP, 5-PPE, and TPP) data points in the tail. This procedure is illustrated in Figure \ref{smooth}.

In Fig. \ref{slopes} we show the slope of the equilibrium dielectric loss at the measuring frequencies of the aging experiment as a function of temperature. The temperature intervals used in the aging experiments are marked with a blue line. For each frequency there is a temperature window where the slope is constant (close to $-1/2$). In this way the measuring frequency and temperature jumps can be fine-tuned such that the proposed method for finding the clock rate is valid. The graphs show that not all measurements are carried out in the optimal regions. The slopes vary a little in the aging temperature interval studied for some of the liquids, and they are not entirely identical to the value above $T_g$. Thus the conditions for the proposed method for determining the clock rate are not fulfilled in all cases. However, one can still obtain data collapse using a slightly incorrect inverse power-law exponent since the error made is the same for all data points. The error simply results in a  vertical shift of the curves in Fig. \ref{kovacsplots} and a horizontal shift in Fig. \ref{tildeR}. A slight variation of the power-law exponent in the measured temperature interval will influence the shape of the master curve and may explain why the data collapse is not perfect.

\acknowledgments 
The centre for viscous liquid dynamics ``Glass and Time'' is sponsored by the Danish National Research Foundation (DNRF).

\end{document}